\documentclass[11pt,a4paper]{scrartcl}

\usepackage{hyperref}

\usepackage[utf8]{inputenc}
\usepackage[T1]{fontenc}
\usepackage{lmodern}

\usepackage{graphicx}
\usepackage{float}
\usepackage{cite}
\usepackage{amssymb}
\usepackage{amsmath}
\usepackage{url}
\usepackage{color}
\usepackage{booktabs}
\usepackage{upgreek}

\usepackage[format=plain,labelfont={bf},font={small}]{caption}

\newcommand{\totd}{\mathrm{d}}
\newcommand{\alphas}{{\alpha_{\mathrm{s}}}}
\newcommand{\Jpsi}{{J/\psi}}
\newcommand{\jpsi}{\Jpsi}
\newcommand{\sps}{\mathrm{SPS}}
\newcommand{\dps}{\mathrm{DPS}}

\newcommand{\refeq}[1]{Eq.\ \eqref{#1}}

\setkomafont{disposition}{\normalcolor\normalfont\bfseries}

\begin{document}

\vspace{-2.0cm}
\begin{flushright}
	MS-TP-16-21
\end{flushright}

\renewcommand*{\thefootnote}{\fnsymbol{footnote}}

\begin{center}
	{\Large \textbf{Double parton scattering in pair-production of $\jpsi$ mesons at the LHC revisited}}\\
	\vspace{.7cm}
	Christoph Borschensky$^{1,}$\footnote{\texttt{christoph.borschensky@uni-tuebingen.de}} and
	Anna Kulesza$^{2,}$\footnote{\texttt{anna.kulesza@uni-muenster.de}}

	\vspace{.3cm}
	\textit{
		$^1$ Institute for Theoretical Physics, University of Tübingen, Auf der Morgenstelle 14, D-72076 Tübingen, Germany\\
		$^2$ Institute for Theoretical Physics, WWU Münster, D-48149 Münster, Germany
	}
\end{center}   

\renewcommand*{\thefootnote}{\arabic{footnote}}
\setcounter{footnote}{0}

\vspace*{0.1cm}
\begin{abstract}
	Double parton scattering (DPS) is studied at the example of $\jpsi$ pair-production in the LHCb and ATLAS experiments of the Large Hadron Collider (LHC) at centre-of-mass energies of $\sqrt{S}=7$, 8, and 13 TeV. We report theoretical predictions delivered to the LHCb and ATLAS collaborations adjusted for the fiducial volumes of the corresponding measurements during Run I and provide new predictions at 13 TeV collision energy. It is shown that DPS can lead to noticeable contributions in the distributions of longitudinal variables of the di-$\jpsi$ system, especially at 13 TeV. The increased DPS rate in double $\jpsi$ production at high energies will open up more possibilities for the separation of single parton scattering (SPS) and DPS contributions in future studies.
\end{abstract}

% #######################
\section{Introduction}
\begin{figure}[t]
	\centering
	\includegraphics[scale=1]{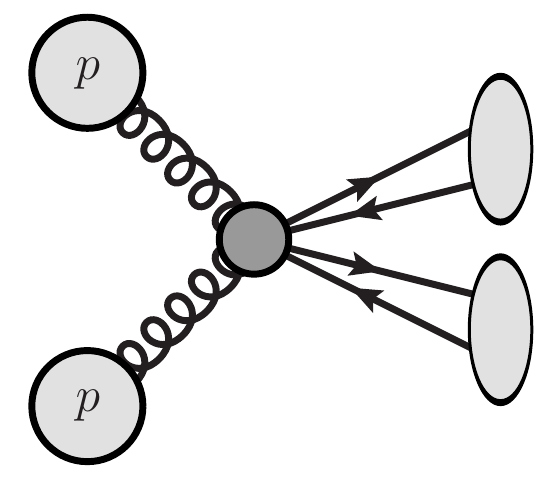}\hspace{2cm}\includegraphics[scale=1]{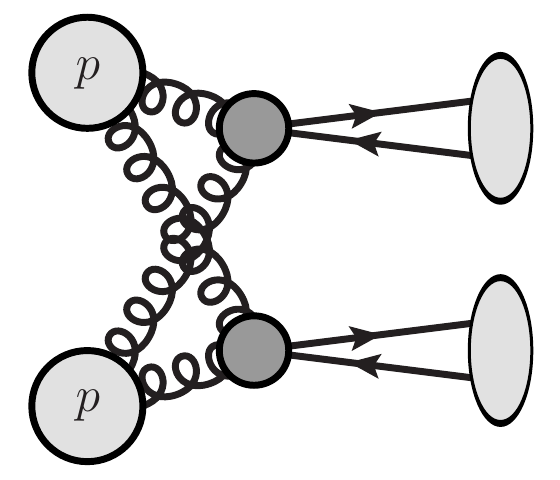}
	\caption{Schematical representation of SPS (left) and DPS (right) for a proton-proton collision. Whereas in the SPS case, the two final-state particles (grey ellipses) originate from the same scattering process (dark grey circle), in the DPS case, two scattering processes of two independent partons from each proton occur.}
	\label{fig:dpsfeyn}
\end{figure}

% #######################
The Large Hadron Collider (LHC) probes collisions of protons at very high energies, resulting in a multitude of final-state particles. With increasing energy, the probability that one hadron-hadron collision leads to more than one scattering process also increases. These additional scattering processes beside the primary hard scattering belong to the group of multi-parton interactions (MPI). Their estimation is important for the correct determination of background from Standard Model processes, for instance when the signal process consists of new physics particles. In particular, double parton scattering (DPS), where two distinct parton interactions arise from the same proton-proton collision, can become likely enough to compete with single parton scattering (SPS) processes, see Fig.~\ref{fig:dpsfeyn}. Therefore, a thorough understanding of these additional contributions is needed for a precise theoretical description of the background at the LHC and will also help to explore the inner structure of protons and nucleons, not being accessible by perturbative calculations.

Double parton scattering has been searched for both in pre-LHC experiments like AFS, UA2, CDF, and D0 as well as by the LHCb and ATLAS collaborations, in 4-jet \cite{Akesson:1986iv,Alitti:1991rd,Abe:1993rv,Aaboud:2016dea}, $\gamma+3$-jet \cite{Abe:1997bp,Abe:1997xk,Abazov:2009gc,Abazov:2014fha},  di-$\gamma+2$-jets \cite{Abazov:2015nnn}, $W+2$-jets \cite{Aad:2013bjm,Chatrchyan:2013xxa}, $\jpsi+W$ \cite{Aad:2014rua}, $\jpsi+Z$ \cite{Aad:2014kba}, open charm \cite{Aaij:2012dz}, $\jpsi$+charm \cite{Aaij:2012dz}, $\Upsilon$+charm \cite{Aaij:2015wpa},  $\jpsi+\Upsilon$ \cite{Abazov:2015fbl} and $\jpsi+\jpsi$ \cite{Abazov:2014qba} final states. On the theoretical side the efforts have concentrated on improving the understanding of the underlying scattering mechanism as well as providing phenomenological predictions. In particular related issues such as correlations and interferences between the two hard scatterings, the role of the perturbative splitting contributions (so-called ``2v1'') and the definition of double parton scattering cross section as well as double parton distributions have been addressed, see e.g.~\cite{Proceedings:2016tff, Bansal:2014paa,Bartalini:2011jp} for a comprehensive review.

A $\jpsi$ pair is a very good candidate to study double parton scattering at the LHC due to relatively high production rates and subsequent decays into muons giving a clear and easily distinguishable signal. Results for the production of $\jpsi$ pairs have been published by LHCb in \cite{Aaij:2011yc}, by D0 in \cite{Abazov:2014qba}, and by CMS in \cite{Khachatryan:2014iia}. Correspondingly, since then there has been a considerable interest to improve theoretical predictions for double $\jpsi$ production both for the SPS and DPS production modes \cite{Kom:2011bd,Baranov:2011ch,Novoselov:2011ff,Kom:2011nu,Gaunt:2011rm,Baranov:2011zz,Martynenko:2012tf,Baranov:2012re,d'Enterria:2013ck,Li:2013csa,Harland-Lang:2014efa,Lansberg:2013qka,Lansberg:2014swa,Sun:2014gca,Mariotto:2015pla,Lansberg:2015lva,He:2015qya,Baranov:2015cle,Shao:2015vga,Likhoded:2016zmk}.

The calculation of conventional single parton scattering contributions to $\jpsi$ pair-production is non-trivial and requires specific methods to account for the non-perturbative mechanisms involved in meson production as well as the short-distance effects. Two widely applied approaches are the colour-singlet model (CSM) \cite{Chang:1979nn,Baier:1981uk,Baier:1983va} and non-relativistic quantum chromodynamics (NRQCD) \cite{Bodwin:1994jh}. In the framework of NRQCD, until not long ago, only the LO predictions for hadronic production in the colour singlet production mode~\cite{Barger:1995vx, Qiao:2002rh, Qiao:2009kg,Berezhnoy:2011xy}, supplemented by the octet corrections~\cite{Li:2009ug,Ko:2010xy}, were known. Recently, the effects of relativistic corrections~\cite{Martynenko:2012tf,Li:2013csa}, NLO corrections and selected NNLO QCD contributions \cite{Lansberg:2013qka,Sun:2014gca, Lansberg:2014swa,Likhoded:2016zmk} as well as an application of the $k_T$ factorisation approach~\cite{Baranov:2011zz,Baranov:2015cle} have been investigated. Additionally, the importance of including contributions from all possible $c\bar{c}$ Fock state configurations relevant for prompt double $\jpsi$ production has been pointed out in~\cite{He:2015qya}. 

This paper documents the predictions of SPS and DPS production of a pair of $\jpsi$, delivered to the LHCb and ATLAS experiments for their ongoing studies of double parton scattering with Run I data.  The work presented here updates the study on $\jpsi$ pair-production reported in \cite{Kom:2011bd}, which in turn was inspired by the first measurement of a double $\jpsi$ signal \cite{Aaij:2011yc}.  Furthermore, predictions for the current LHC run at a centre-of-mass energy of $\sqrt{S}=13$~TeV are provided. We also perform a comparison with CMS data \cite{Khachatryan:2014iia} and more thoroughly with theoretical predictions for double $\jpsi$ production obtained by another group \cite{Lansberg:2014swa}.

The outline is as follows. In section~\ref{sec:theo_setup}, the theoretical setup of \cite{Kom:2011bd, Kom:2011nu} used for both the SPS and DPS cross section calculations is reviewed, followed by a listing of Monte Carlo parameters for event simulation in section~\ref{sec:monte_sim}. We present numerical results for total cross sections and kinematic distributions for a choice of experimentally accessible variables in section~\ref{sec:kin_dis}. At last, we conclude in section~\ref{sec:conclusions}.

% #######################
\section{Theoretical setup}\label{sec:theo_setup}
% #######################
% #######################
\subsection{Single parton scattering}\label{subsec:spsfact}
% #######################

In this work, the SPS contributions will be considered utilising a leading-order (LO) colour-singlet result presented in \cite{Li:2009ug} and including radiative corrections from parton showering. The details of the implementation are described in section~\ref{sec:monte_sim} and the SPS results obtained in this way are compared to the NLO calculations of~\cite{Lansberg:2014swa} in section~\ref{sec:complansberg}. 

As it was pointed out in \cite{Lansberg:2014swa}, the prompt production of $\jpsi$ mesons comprises feed-down from the decay of $\chi_c$ and $\psi'$ at a non-negligible amount of roughly 85\%. The SPS calculation of \cite{Li:2009ug} is for direct production of $\jpsi$ pairs only, so in the following, all SPS cross sections will be considered for prompt production, $\sigma^{\mathrm{prompt}} = 1.85\times\sigma^{\mathrm{direct}}$. The DPS results implicitely include feed-down contributions due to the fit to experimental data.

To include some higher-order effects in our SPS predictions, in addition to using NLO PDFs, we enable initial-state radiation or parton showering within the \texttt{Herwig} \cite{Bahr:2008pv,Bellm:2015jjp} framework. Furthermore, if denoted, we also add effects of intrinsic transverse momentum of the initial-state partons using a Gaussian model in \texttt{Herwig} with a root mean square $\sigma_{p_T}$ of 2~GeV. We have checked that the predictions do not depend strongly on the actual numerical value, and it will be seen in the following sections that the effect of the intrinsic transverse momentum is rather mild on the distributions.

% #######################
\subsection{Factorisation approach for double parton scattering}\label{subsec:dpsfact}
% #######################
DPS production of a $\jpsi$-pair is described using an approximation in which the DPS cross section factorises into a product of two hard-scattering cross sections describing single-$\jpsi$ production which are independent from each other:
\begin{align}
	\totd\sigma^{2\jpsi}_\dps = \frac{\totd\sigma^{\jpsi}_\sps\,\totd\sigma^{\jpsi}_\sps}{2\sigma_{\mathrm{eff}}}\label{dpsfact}.
\end{align}
This customary approximation assumes factorization of the transverse and longitudinal components in the generalized parton distribution function. We refer the reader to~\cite{Diehl:2011tt,Diehl:2011yj,Diehl:2015bca,Gaunt:2011xd,Gaunt:2012dd,Gaunt:2014ska,Blok:2011bu,Blok:2013bpa,Manohar:2012jr,Manohar:2012pe} for a discussion of the validity of the approximation and the status of understanding factorization in DPS.
The SPS cross section for single-$\jpsi$ production is given as
\begin{align}
	\totd\sigma^{\jpsi}_\sps = \sum_{a,b}f_a(x_1,\mu_F^2)\,f_b(x_2,\mu_F^2)\,\totd\hat{\sigma}^{\jpsi}_\sps\,\totd x_1\,\totd x_2
\end{align}
with a sum over the initial-state flavours $a,b$ and the parton distribution functions $f(x,\mu_F^2)$ and
\begin{align}
	\totd\hat{\sigma}^{\jpsi}_\sps = \frac{1}{2\hat{s}}\overline{|\mathcal{M}_{ab\to\jpsi+X}|^2}\,\totd\mathrm{PS}_{\jpsi+X}\label{xssingle}
\end{align}
the partonic cross section for single $\jpsi$ production with the corresponding matrix elements $\mathcal{M}_{ab\to\jpsi+X}$ and the phase space $\totd\mathrm{PS}_{\jpsi+X}$. $X$ denotes any additional final state which is not a $\jpsi$, and therefore not of interest for $\jpsi$ production. The factor $\sigma_{\mathrm{eff}}$ is assumed to only depend on the transverse structure of the proton, and should therefore be process and energy independent if the factorisation of \refeq{dpsfact} holds. It is the main quantity to be extracted by a DPS experiment. 

Theoretical description of single quarkonium production~\cite{Campbell:2007ws,Gong:2008ft,Butenschoen:2010rq,Ma:2010yw,Khoze:2004eu,Artoisenet:2007xi,Artoisenet:2008fc,Brodsky:2009cf,Butenschoen:2011yh,Gong:2012ug,Chao:2012iv} is challenging even within the NRQCD framework \cite{Brambilla:2014jmp}. Given that LHCb can trigger over low $p_T$ muons it is important to describe the low $p_T$ production accurately.  Therefore in this work we choose to model the low $p_T$ region and use the same setup as in \cite{Kom:2011bd}. It relies on the matrix element as in \refeq{xssingle} given by:
\begin{align}
	\overline{|\mathcal{M}_{ab\to\jpsi+X}|^2} =
	\begin{cases}
		K\exp\left(-\kappa\frac{p_T^2}{m_\jpsi^2}\right) & \text{for } p_T \leq \langle p_T\rangle\\
		K\exp\left(-\kappa\frac{\langle p_T\rangle^2}{m_\jpsi^2}\right)\left[1+\frac{\kappa}{n}\frac{p_T^2-\langle p_T\rangle^2}{m_\jpsi^2}\right]^{-n} & \text{for } p_T > \langle p_T\rangle,
	\end{cases}\label{crystalball}
\end{align}
which describes a fit of data from the LHCb \cite{Aaij:2011jh}, ATLAS \cite{Aad:2011sp}, CMS \cite{Khachatryan:2010yr}, and CDF \cite{Acosta:2004yw} experiments to a Crystal Ball function. In \refeq{crystalball}, $K = \lambda^2\kappa\hat{s}/m_\jpsi^2$, and the fit parameters are determined to be $\kappa = 0.6$ and $\lambda = 0.327$ for $n = 2$ and $\langle p_T\rangle = 4.5$~GeV. While the work of \cite{Lansberg:2014swa} updated the fit parameters to include more recent measurements of single $\jpsi$ production, we have checked that the change in predictions for DPS production is only moderate and well within the uncertainty on $\sigma_{\mathrm{eff}}$. We have also checked that for the available measurement of single $\jpsi$ production at 13 TeV from the LHCb experiment \cite{Aaij:2015rla} with $\sigma^{\jpsi}_{\mathrm{exp}} = 15.30 \pm 0.03 \pm 0.86$ $\upmu$b, the fit parameters still produce results at 13 TeV consistent with the LHCb measurement, $\sigma^{\jpsi}_{\mathrm{fit}} = 15.83$ $\upmu$b.

% #######################
\subsection{Details of simulation}\label{sec:monte_sim}
% #######################
The public Monte Carlo event generator \texttt{Herwig-7.0.3} \cite{Bahr:2008pv,Bellm:2015jjp} has been used to simulate double $\jpsi$ production at the LHC via SPS. The central values for the renormalisation and factorisation scales are chosen as the transverse mass of a single $\jpsi$, $\mu_R = \mu_F = m_T = \sqrt{m_\jpsi^2+p_T^2}$ with the physical $\jpsi$ mass $m_\jpsi = 3.097$~GeV \cite{Agashe:2014kda}. One parameter appearing in the calculation of the SPS cross section of \cite{Li:2009ug} is the charm quark mass which we set to $m_c = \frac{1}{2}m_\jpsi$ which corresponds to the LO choice of the hadron mass in a NRQCD calculation \cite{Ko:2010xy}. Another input parameter entering the SPS calculation is the non-perturbative wave function of the $\jpsi$ meson at the origin. In the following computations, it is set to $|R(0)|^2 = 0.92$ GeV$^3$ \cite{Bodwin:2007fz, Butenschoen:2011yh}. It should be noted that a variation of this parameter can be achieved by multiplying the SPS cross section by a factor of $\left(|R(0)_{\mathrm{new}}|^2/|R(0)|^2\right)^2$, where $|R(0)_{\mathrm{new}}|^2$ is the new value of the wave function. 

We use MSTW2008 NLO parton distribution functions \cite{Martin:2009iq} for the SPS predictions including initial-state radiation and for DPS. The parton distribution functions are accessed via the LHAPDF 6 library \cite{Buckley:2014ana}. The $\jpsi$ mesons are assumed to decay isotropically into a pair of opposite-sign (OS) muons with a branching ratio of $\mathrm{BR}(\jpsi\to 2\mu) = 0.05935$. Out of the two possible combinations of choosing OS muon pairs, the one with an invariant mass closest to $m_\jpsi$ is chosen. From these pairs, properties of the $\jpsi$ are reconstructed. To optimise the data samples collected by the experiments for a DPS analysis, a certain set of cuts on transverse momentum and (pseudo-)rapidity of the $\jpsi$ as well as their decay products is applied.

Unless otherwise specified, we use the CDF value of $\sigma_{\mathrm{eff}} = 14.5\pm 1.7^{+1.7}_{-2.3}$~mb \cite{Abe:1997xk}. A more recent double-$\jpsi$ study by D0 reports a lower value of $4.8\pm 0.5\pm 2.5$~mb \cite{Abazov:2014qba}, but given that most of other experiments measure  higher values, see e.g.~\cite{Aaboud:2016dea}, and the difficulty of theoretical modelling of $\sigma_{\mathrm{eff}}$, we choose the CDF value. With its relatively wide error bars it then accounts to a large extent for the observed span in values of  $\sigma_{\mathrm{eff}} $. We also note that this value is in accordance with the phenomenological estimates \cite{Blok:2013bpa, Gaunt:2014rua} taking into account in our framework (Eq.~\ref{dpsfact}) the so-called ``2v2'' and ``2v1'' contributions, i.e.\ contributions from two separate parton ladders or from one ladder and another ladder created by a perturbative splitting of a single parton, respectively. As found out in~\cite{Gaunt:2014rua}, the shapes of the transverse momentum and rapidity distributions for the two types of production mechanisms remain very similar, justifying our effective approach of considering only the conventional 2v2 scattering.

In a similar manner as for the non-perturbative wave function at the origin, the DPS results for a different value of $\sigma_{\mathrm{eff}}$ can be obtained by rescaling our DPS cross section with a factor of $\sigma_{\mathrm{eff}}/\sigma_{\mathrm{eff,new}}$. The DPS predictions have been cross checked with two independent numerical in-house implementations.

% #######################
\subsubsection{LHCb cuts}\label{subsec:lhcbcuts}
% #######################
The LHCb experiment, being a forward spectrometer, mainly selects events in the forward-scattering region with low transverse momentum. The cuts are\footnote{From private communications with Andrew Cook.}:
\begin{itemize}
	\item $p_{T,\jpsi} < 10$ GeV,
	\item $2 < y_\jpsi < 4.2$,
\end{itemize}
with $p_{T,\jpsi}$ being the transverse momentum and $y_\jpsi$ the rapidity of a single $\jpsi$. Any further cuts on the muons as the decay products of the di-$\jpsi$ are not relevant for the theoretical predictions presented in this work, as they are already taken into account in the efficiency correction of the data.

% #######################
\subsubsection{ATLAS cuts}\label{subsec:atlascuts}
% #######################
The ATLAS experiment probes the $\jpsi$ in the central region imposing a minimum transverse momentum and a more central rapidity region. Additionally, several cuts are also applied to the muons\footnote{From private communications with Ben Weinert.}:
\begin{itemize}
	\item[1)] $p_{T,\jpsi} > 8.5$ GeV,
	\item[2)] $|y_\jpsi| < 2.1$,
	\item[3)] $p_{T,\mu} > 2.5$ GeV,
	\item[4)] $|\eta_\mu| < 2.3$,
	\item[5)] at least 1 $\jpsi$ with both $p_{T,\mu} > 4$ GeV,
\end{itemize}
with $p_{T,\mu}$ being the transverse momentum and $\eta_\mu$ the pseudorapidity of one muon.

% #######################
\section{Cross sections and kinematic distributions}\label{sec:kin_dis}
% #######################
\begin{table}[t]
	\centering
	\begin{tabular}{ccc}
		\toprule
		$\sigma$ (pb)  & PS+$\sigma$(2 GeV) & DPS\\
		\midrule
		LHCb ($\sqrt{S}=$ 7 TeV) & $11.4^{-7.6}_{+0.5}$ & $12.0^{+4.6}_{-2.3}$\\
		ATLAS ($\sqrt{S}=$ 7 TeV) & $\left(4.71^{-3.16}_{+0.21}\right)\times 10^{-3}$ & $\left(3.98^{+1.51}_{-0.75}\right)\times 10^{-3}$\\
		\midrule
		LHCb ($\sqrt{S}=$ 8 TeV) &  $12.5^{-8.8}_{+0.9}$ & $14.2^{+5.4}_{-2.7}$\\
		ATLAS ($\sqrt{S}=$ 8 TeV) & $\left(5.24^{-3.68}_{+0.37}\right)\times 10^{-3}$ & $\left(5.08^{+1.93}_{-0.96}\right)\times 10^{-3}$\\
		\midrule
		LHCb ($\sqrt{S}=$ 13 TeV) & $16.5^{-13.4}_{+2.9}$ & $25.2^{+9.6}_{-4.8}$\\
		ATLAS ($\sqrt{S}=$ 13 TeV) & $\left(8.09^{-6.58}_{+1.43}\right)\times 10^{-3}$ & $\left(11.9^{+4.5}_{-2.3}\right)\times 10^{-3}$\\
		\bottomrule
	\end{tabular}
	\caption{Total cross sections for SPS and DPS production of a $\jpsi$ pair for different centre-of-mass energies and cuts.  PS+$\sigma$(2 GeV) denotes the addition of initial-state radiation and intrinsic transverse momentum to the LO calculation in \texttt{Herwig}. All numbers include the branching ratio factor of $\mathrm{BR}^2(\jpsi\to 2\mu)$. The uncertainties on the PS+$\sigma$(2 GeV) numbers correspond to a simultaneous variation of the renormalisation and factorisation scales up and down by a factor of two, while the uncertainty on the DPS numbers corresponds to the uncertainty of our value of $\sigma_{\mathrm{eff}}$ used, see section~\ref{sec:monte_sim}.}
	\label{tbl:totxsectspsdps}
\end{table}
\begin{figure}[h!]
	\centering
	\begin{tabular}{cc}
		\includegraphics[width=0.47\columnwidth]{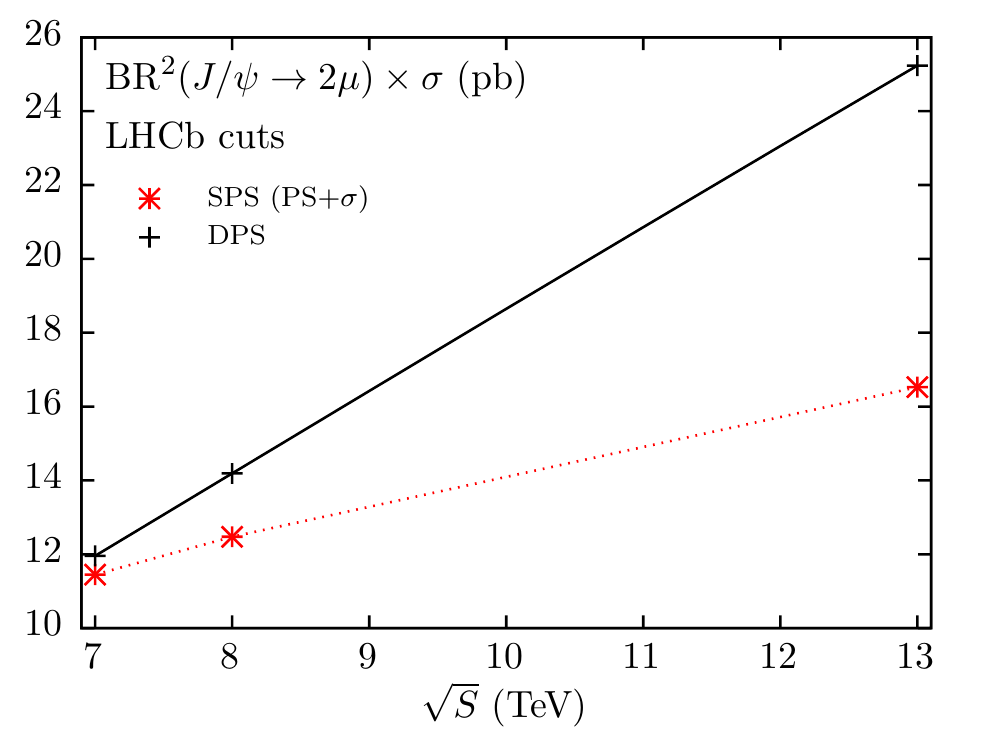}&
		\includegraphics[width=0.47\columnwidth]{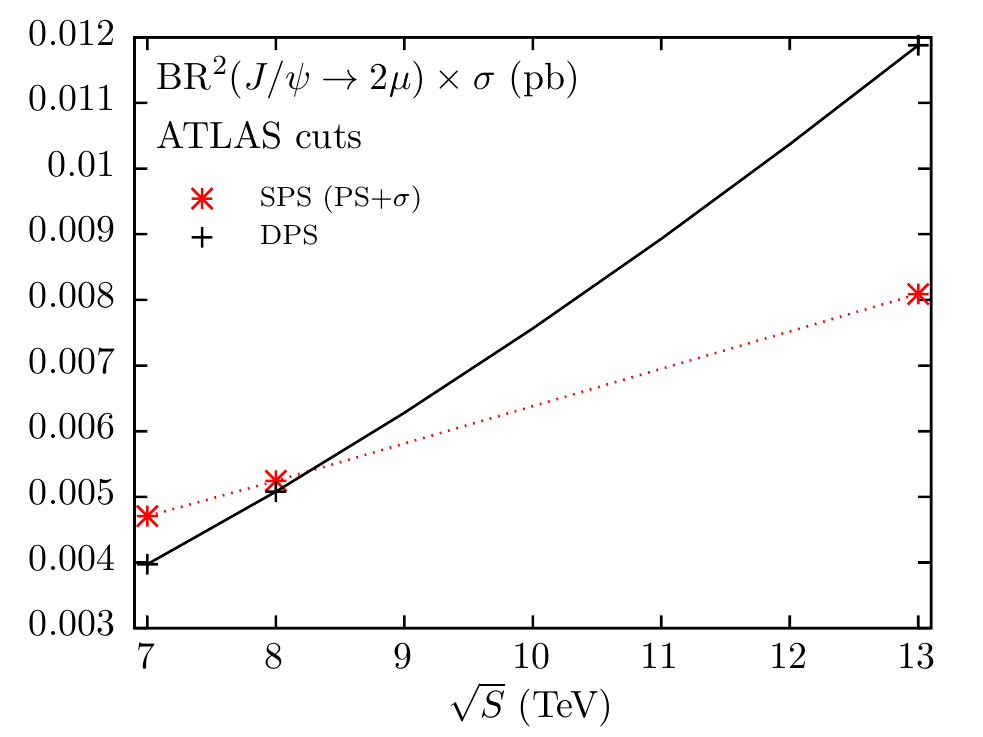}\\
		(a) & (b)
	\end{tabular}
	\caption{Total cross section dependence of the centre-of-mass energy of the collider for the LHCb (a) and the ATLAS cuts (b). Shown are the SPS prediction with parton showering and intrinsic transverse momentum of the partons, and the DPS contribution. The lines for SPS are interpolated between the points.}
	\label{fig:totxsec}
\end{figure}
Table~\ref{tbl:totxsectspsdps} and Fig.~\ref{fig:totxsec} show the total cross sections for SPS and DPS production of $\jpsi$ pairs with the LHCb and the ATLAS cuts at different centre-of-mass energies returned by our simulation. From Fig.~\ref{fig:totxsec}, it can be clearly seen that, while at 7 and 8 TeV, the SPS and DPS contributions are of roughly similar size, the DPS contributions grow more rapidly with increasing collider energy, and are larger by almost a factor of 2 with respect to the SPS cross sections at 13 TeV. This is important for the current LHC run, as it shows that DPS processes will occur more prevalently and that they will be more easily distinguishable from SPS.

Another interesting observation is the large difference in magnitude of the LHCb and ATLAS values, being smaller by over three orders of magnitude in the case of the ATLAS predictions. We checked that mainly cut no.\ 5 of section~\ref{subsec:atlascuts}, requiring both muons that stem from the same decay of one $\jpsi$ to have $p_{T,\mu} > 4$ GeV, and to a lesser amount cut no.\ 1, requiring each $\jpsi$ to have a $p_{T,\jpsi} > 8.5$ GeV, lower the cross section values significantly. While we expect that due to the upper $p_T$ cut on each $\jpsi$, the total cross section predictions for LHCb are only little affected by the higher order QCD corrections and the colour octet contributions not present in the CSM, we caution the same cannot hold to the same extent for the ATLAS predictions, probing much larger $p_T$ values of the $\jpsi$. In consequence, it is reasonable to assume that the ATLAS predictions in table~\ref{tbl:totxsectspsdps} underestimate the total cross section.

% #######################
\subsection{LHCb predictions}\label{sub:lhcb_predictions}
% #######################
\begin{figure}[t]
	\centering
	\begin{tabular}{cc}
		\includegraphics[width=0.47\columnwidth]{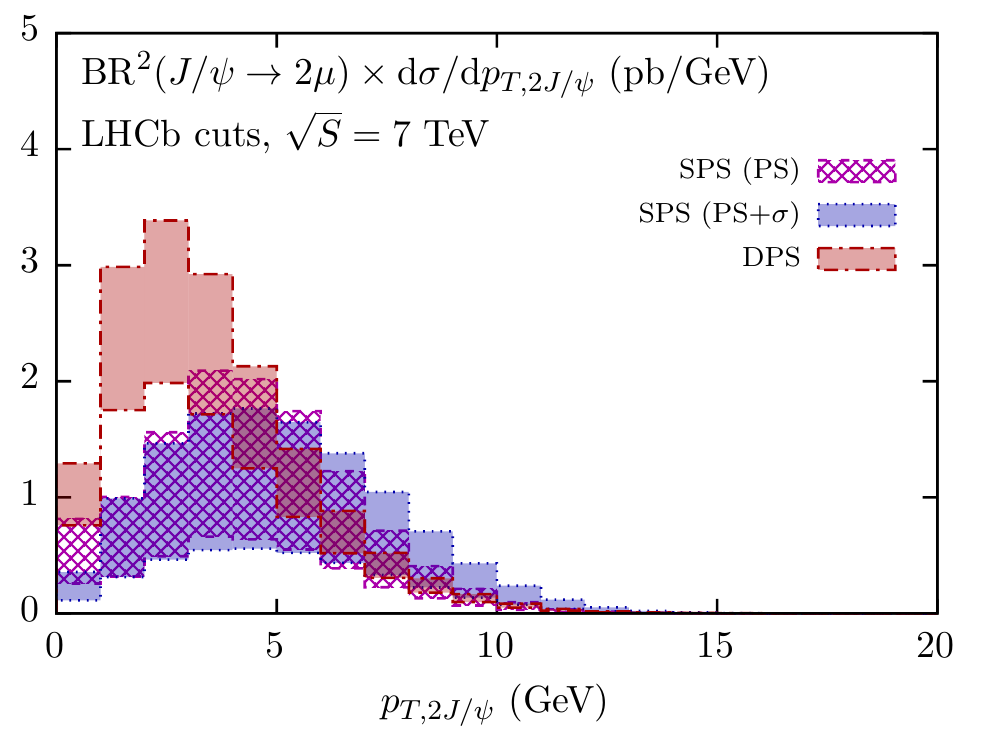}&
		\includegraphics[width=0.47\columnwidth]{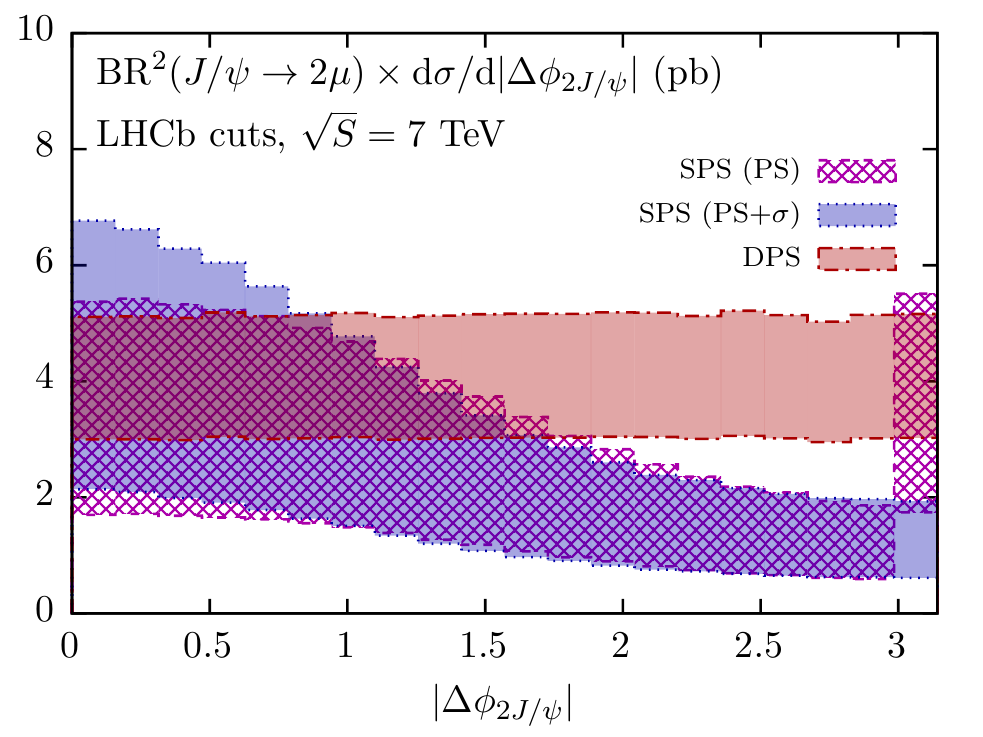}\\
		(a) & (b)\\
		\includegraphics[width=0.47\columnwidth]{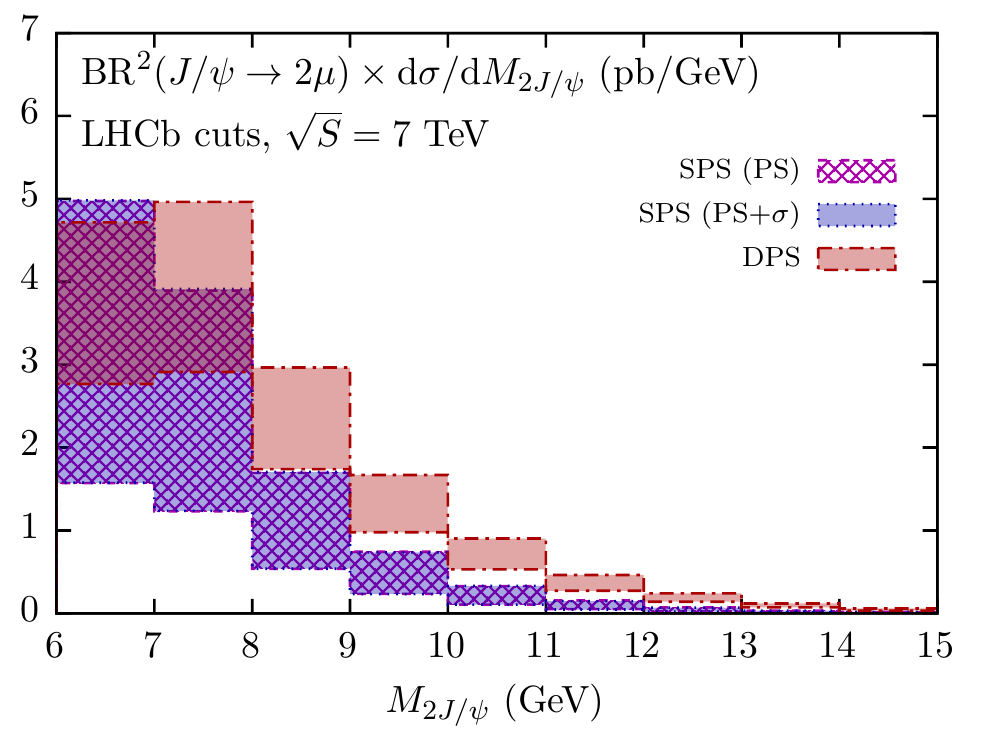}&
		\includegraphics[width=0.47\columnwidth]{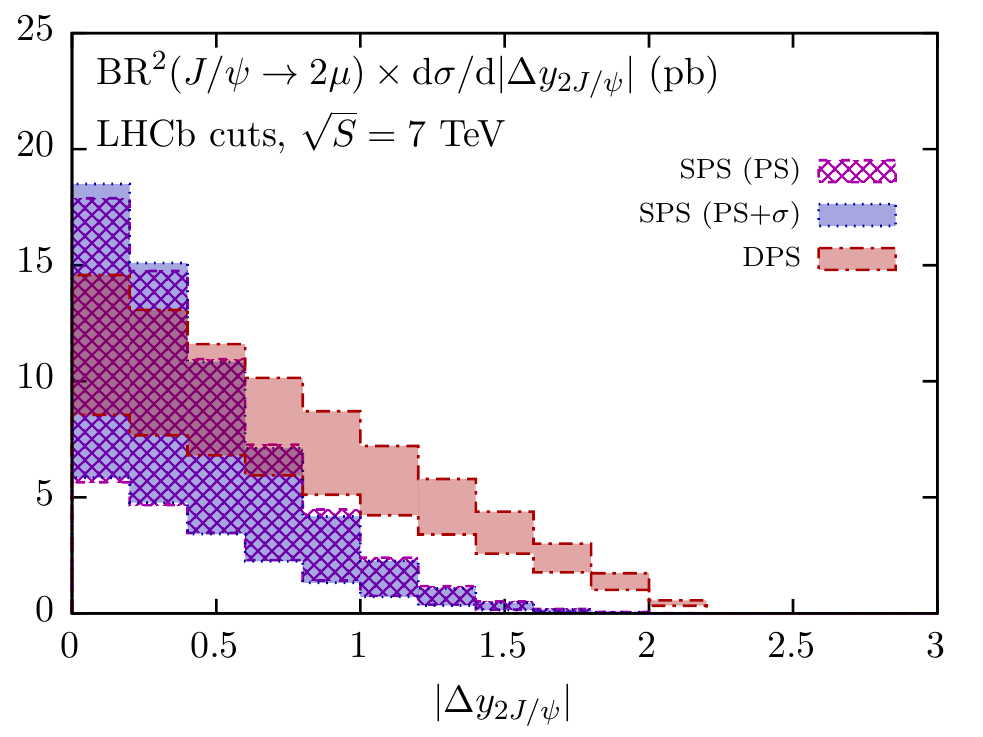}\\
		(c) & (d)
	\end{tabular}
	\caption{Differential distributions for the LHCb cuts at a collider energy of $\sqrt{S} = 7$~TeV. Shown are the transverse momentum of the di-$\jpsi$ system (a), the azimuthal angular separation (b), the invariant mass spectrum (c), and the rapidity separation (d) for various SPS and DPS predictions, see the accompanying text for an explanation of the lines.}
	\label{fig:lhcb_plots}
\end{figure}
The LHCb collaboration made a first measurements of double-$\jpsi$ production over four years ago with data collected at a centre-of-mass energy of $\sqrt{S} = 7$ TeV \cite{Aaij:2011yc}. They found a total cross section of $\sigma^{2\jpsi}_{\mathrm{exp}} = 5.1 \pm 1.0 \pm 1.1$~nb in the fiducial volume defined by the rapidity range $2 < y_\jpsi < 4.5$ and a maximum transverse momentum $p_{T,\jpsi} < 10$~GeV. Despite the slightly changed fiducial volume in regard to the rapidity of the $\jpsi$, our result for SPS with parton showering and intrinsic transverse momentum added to DPS, $\mathrm{BR}^2(\jpsi\to 2\mu)\times\sigma^{2\jpsi}_{\mathrm{SPS+DPS}} = 23.4^{+4.6}_{-7.9}$~pb (see table~\ref{tbl:totxsectspsdps}), agrees with the experimental measurement multiplied by the squared branching ratio $\mathrm{BR}^2(\jpsi\to 2\mu)\times\sigma^{2\jpsi}_{\mathrm{exp}} = 18.0 \pm 3.5 \pm 3.9$~pb within the experimental uncertainties. The previous theoretical study in \cite{Kom:2011bd} analysed these results and compared the predicted invariant mass distribution to the one measured by LHCb  \cite{Aaij:2011yc}. It was also proposed there \cite{Kom:2011bd} to separate the DPS component from the data by using correlations in the longitudinal direction of the $\jpsi$ and imposing constraints on a minimum rapidity separation. The method has already been used in the experimental analysis of double $\jpsi$ production by D0 \cite{Abazov:2014qba}. In this section, we analyse the updated fiducial volume as defined by the cuts in section~\ref{subsec:lhcbcuts}.

In Fig.~\ref{fig:lhcb_plots}, a selection of four differential distributions is shown: transverse momentum of the di-$\jpsi$ system $p_{T,2\jpsi}$, azimuthal angular separation $|\Delta\phi_{2\jpsi}|$, invariant mass $M_{2\jpsi}$, and rapidity separation $|\Delta y_{2\jpsi}|$. Shown are predictions for SPS production including only parton showering/initial-state radiation (PS) and both initial-state radiation as well as intrinsic transverse momentum of the initial-state partons (PS$+\sigma$). The error band for the SPS predictions is calculated from a simultaneous variation of the renormalisation and factorisation scales up and down by a factor of 2 with respect to the central value, $\mu_R = \mu_F = k_\mu m_T$, $k_\mu = (0.5,1,2)$ with $m_T$ given in section~\ref{sec:monte_sim}. The DPS predictions are calculated as outlined in section~\ref{subsec:dpsfact}, with the error band now corresponding to the uncertainties on the effective cross section $\sigma_{\mathrm{eff}}$ given in section~\ref{sec:monte_sim}.

It can be seen that the DPS contribution is comparable in size to the SPS background and therefore makes up a significant part of the sum of signal\footnote{We remark that ``signal'' refers to the DPS predictions.} and background predictions. In particular, for the $p_T$ spectra of the di-$\jpsi$ system, Fig.~\ref{fig:lhcb_plots}~(a), the DPS contribution makes up the major part at low $p_{T,2\jpsi}$. However, besides the similar size of SPS and DPS contributions, it is difficult to differentiate the DPS signal from the background in terms of the shape of this distribution. The azimuthal angular separation between the two $\jpsi$ mesons, shown in Fig.~\ref{fig:lhcb_plots}~(b), is not very well suited for this distinction either, as the DPS signal shows a uniform distribution in the transverse direction, which, although strongly differing from the pure SPS LO result only allowing back-to-back scattering (not shown here, see \cite{Kom:2011nu} for the corresponding histogram), is made less significant by higher-order radiation strongly distorting the SPS background, even peaking towards low or no angular separation at all. It can be seen in Figs.~\ref{fig:lhcb_plots}~(a) and (b) that the inclusion of intrinsic transverse momentum of the initial-state partons in addition to parton showering leads to a further smearing out of the distributions. As it was noted in \cite{Kom:2011nu}, an analysis of the longitudinal components such as the invariant mass, see Fig.~\ref{fig:lhcb_plots}~(c), or the rapidity separation between the two $\jpsi$ in Fig.~\ref{fig:lhcb_plots}~(d) leads to more conclusive results, because at a high rapidity separation, the full signal+background predictions are primarily made up by the DPS signal with almost no SPS background expected.

The new fiducial volume with a slightly lower maximum rapidity of a $\jpsi$ does not change the results significantly compared to the ones in \cite{Kom:2011nu}. We note that, while in \cite{Kom:2011nu}, also muon cuts have been taken into account, our results do not include muon cuts. Also, we study here prompt $\jpsi$ production, what corresponds to increasing the SPS numbers by a factor of 1.85 with respect to direct $\jpsi$ production. Thus, while the absolute values of the cross sections differ, the shapes of the distributions stay approximately the same.

% #######################
\subsubsection{Predictions at 13 TeV}
% #######################
\begin{figure}[t]
	\centering
	\begin{tabular}{cc}
		\includegraphics[width=0.47\columnwidth]{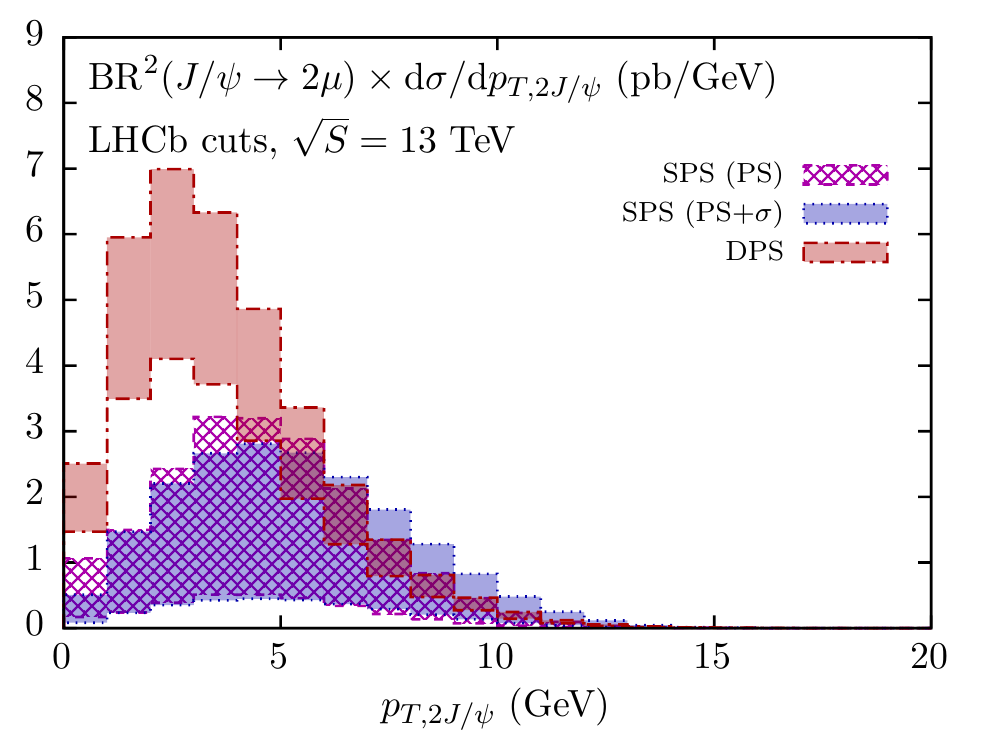}&
		\includegraphics[width=0.47\columnwidth]{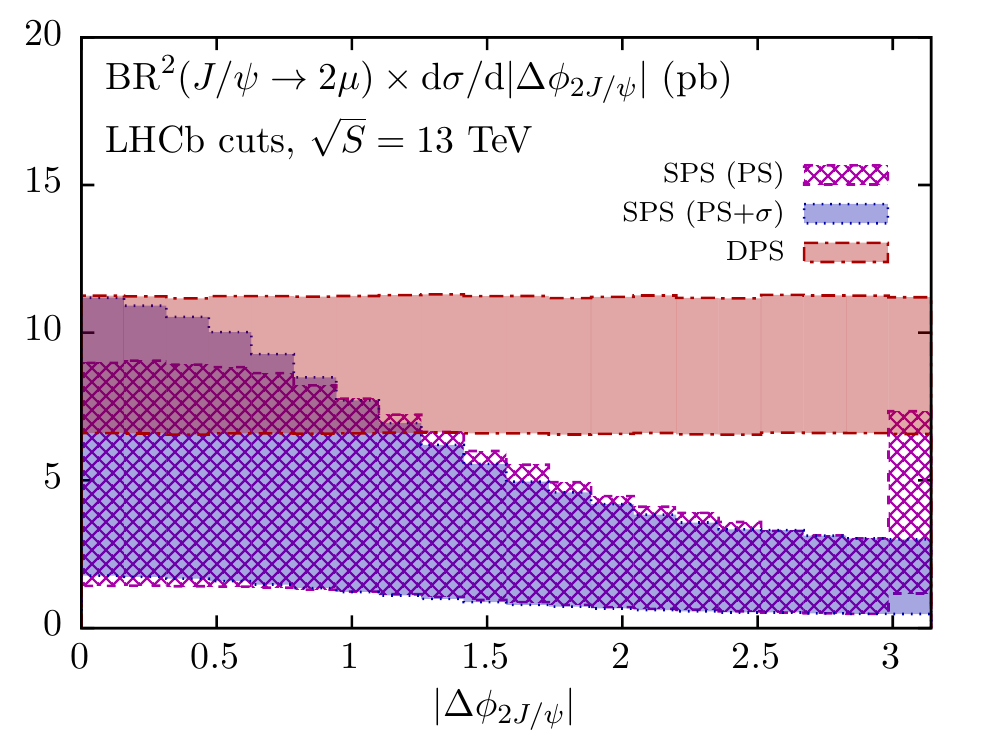}\\
		(a) & (b)\\
		\includegraphics[width=0.47\columnwidth]{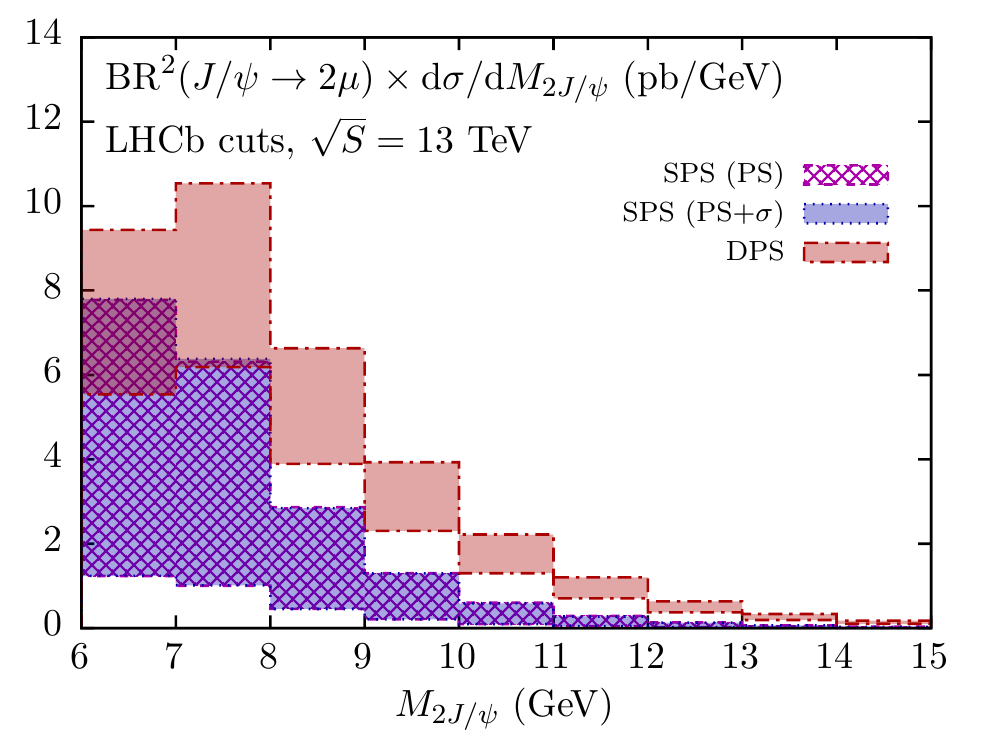}&
		\includegraphics[width=0.47\columnwidth]{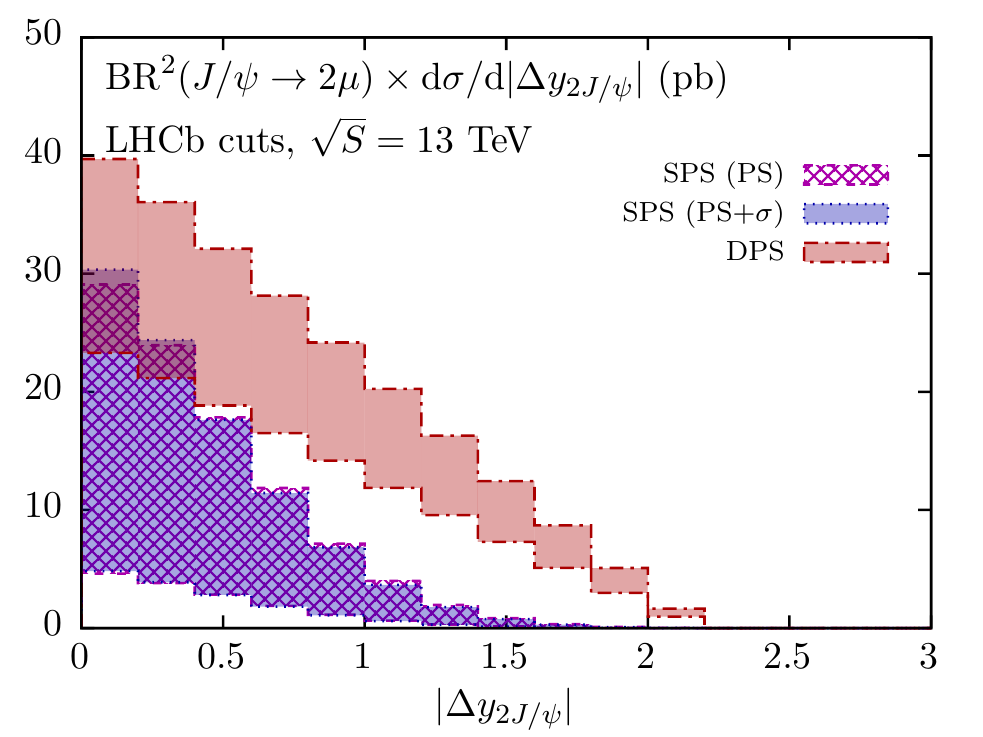}\\
		(c) & (d)
	\end{tabular}
	\caption{Differential distributions for the LHCb cuts at a collider energy of $\sqrt{S} = 13$ TeV. The distributions are the same as for Fig.~\ref{fig:lhcb_plots}.}
	\label{fig:lhcb_plots13}
\end{figure}
At a collider energy of 13 TeV, shown in Fig.~\ref{fig:lhcb_plots13}, we see that the DPS contributions now dominate over SPS in almost all of the bins in the distributions. While the azimuthal angular separation of Fig.~\ref{fig:lhcb_plots13}~(b) still does not serve well to distinguish between DPS and SPS, the lower $p_{T,2\jpsi}$ range of Fig.~\ref{fig:lhcb_plots13}~(a), the higher $M_{2\jpsi}$ region of Fig.~\ref{fig:lhcb_plots13}~(c), and the higher $|\Delta y_{T,2\jpsi}|$ range of Fig.~\ref{fig:lhcb_plots13}~(d) are now dominated much more strongly by DPS. We stress, however, that the $p_{T,2\jpsi}$ distribution is very susceptible to higher-order corrections which can change the shape significantly (at LO, the $p_{T,2\jpsi}$ distribution is given as an infinitely high peak at $p_{T,2\jpsi} = 0$~GeV). In the low $p_T$ regime, though, if we do not use a full NLO calculation for our predictions, we expect that the $p_T$ spectrum should be well described by a parton shower Monte Carlo. The invariant mass and rapidity distributions are more stable with respect to the higher-order corrections, as will also be seen later in section~\ref{sec:complansberg}.

% #######################
\subsection{ATLAS predictions}
% #######################
\begin{figure}[t]
	\centering
	\begin{tabular}{cc}
		\includegraphics[width=0.47\columnwidth]{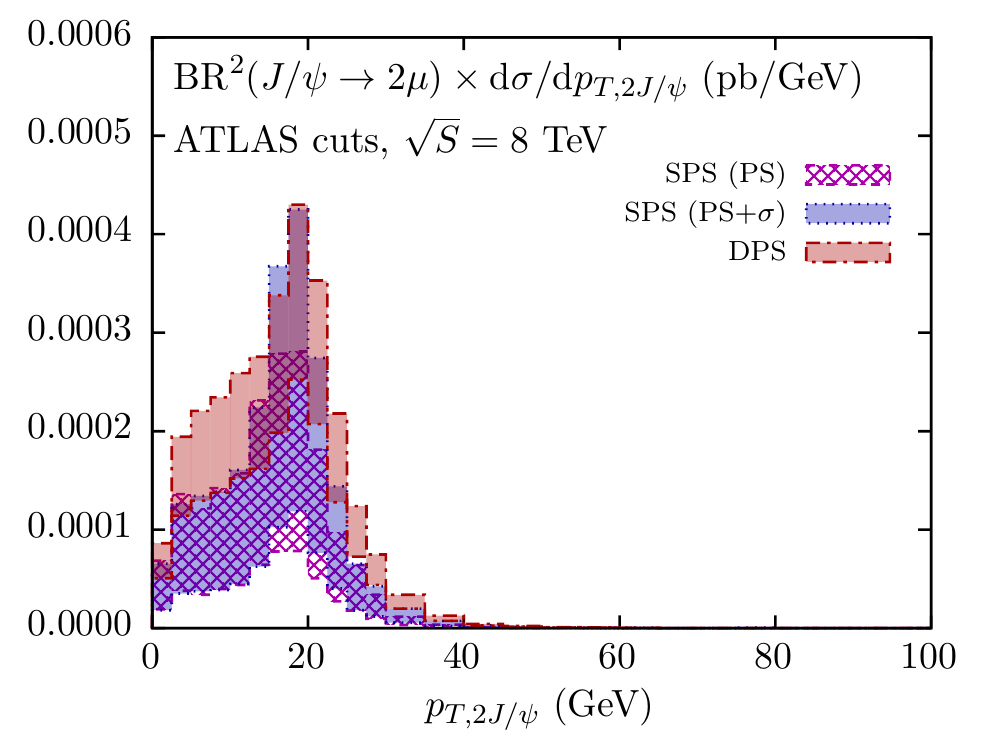}&
		\includegraphics[width=0.47\columnwidth]{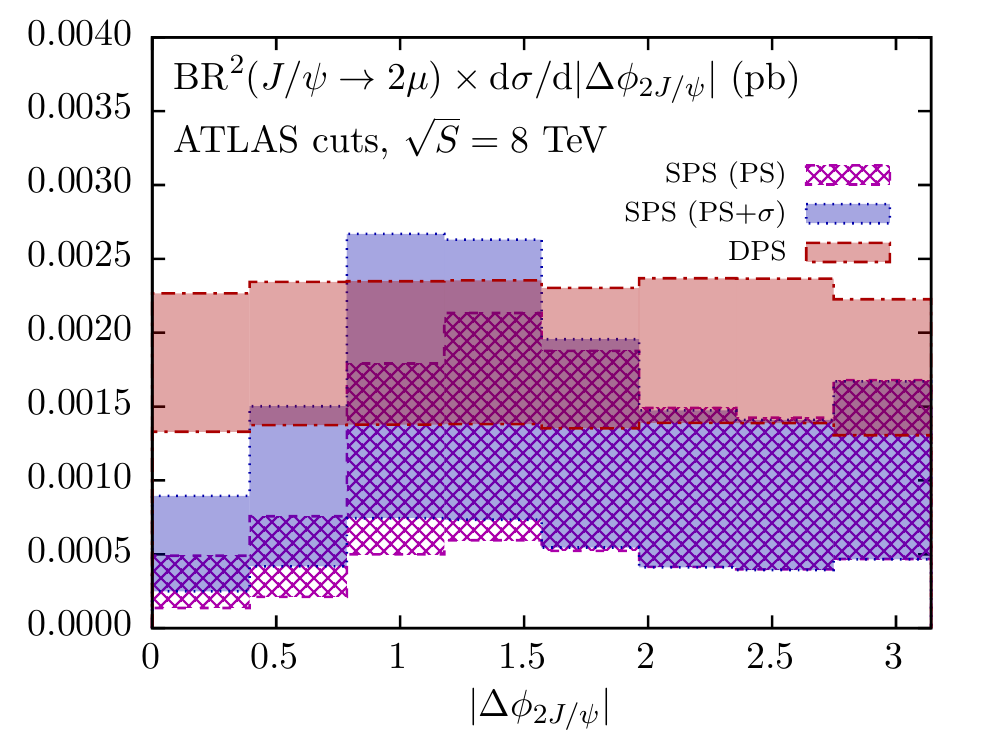}\\
		(a) & (b)\\
		\includegraphics[width=0.47\columnwidth]{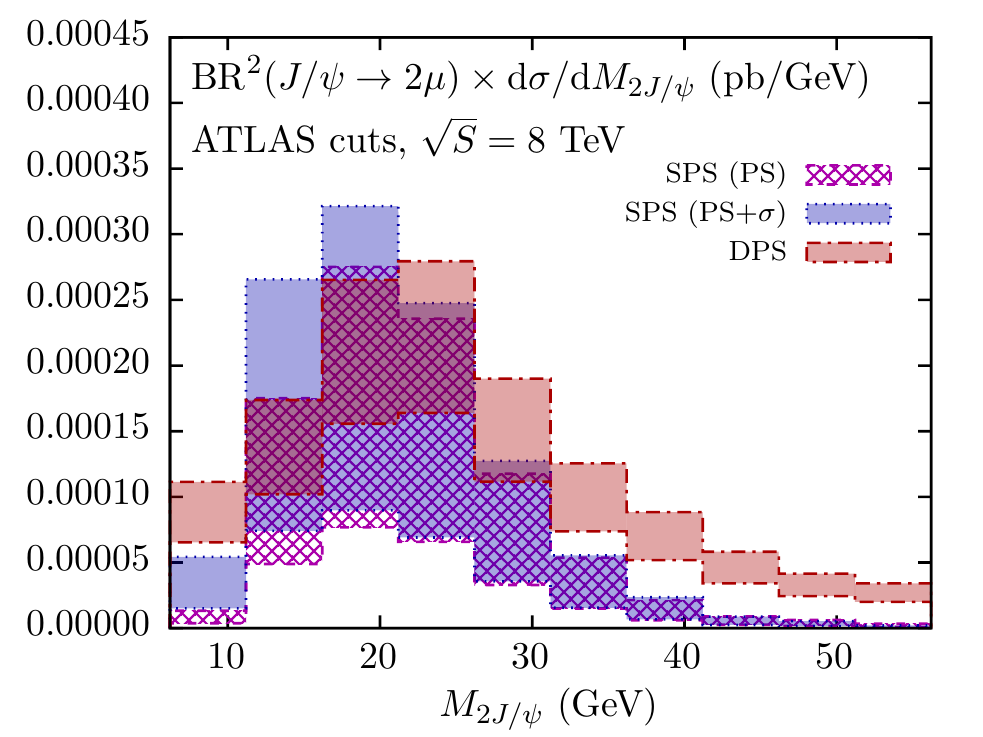}&
		\includegraphics[width=0.47\columnwidth]{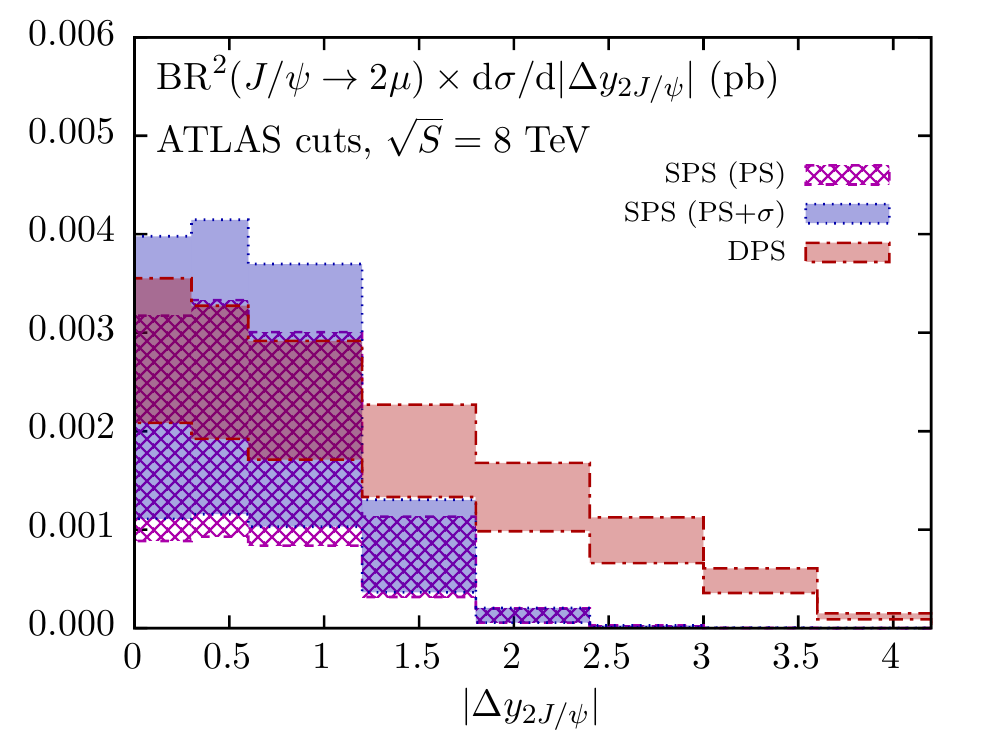}\\
		(c) & (d)
	\end{tabular}
	\caption{Differential distributions for the ATLAS cuts at a collider energy of $\sqrt{S} = 8$ TeV. The distributions are the same as for Fig.~\ref{fig:lhcb_plots}.}
	\label{fig:atlas_plots}
\end{figure}
In this subsection we show predictions for a collider energy of  $\sqrt{S} = 8$ TeV and ATLAS cuts. They mainly differ from the LHCb ones by imposing a minimum transverse momentum for each $\jpsi$, effectively probing the high-$p_T$ region unlike the LHCb experiment. Another obvious difference concerns rapidity regions which both detectors probe. 

\begin{figure}[t]
	\centering
	\begin{tabular}{cc}
		\includegraphics[width=0.47\columnwidth]{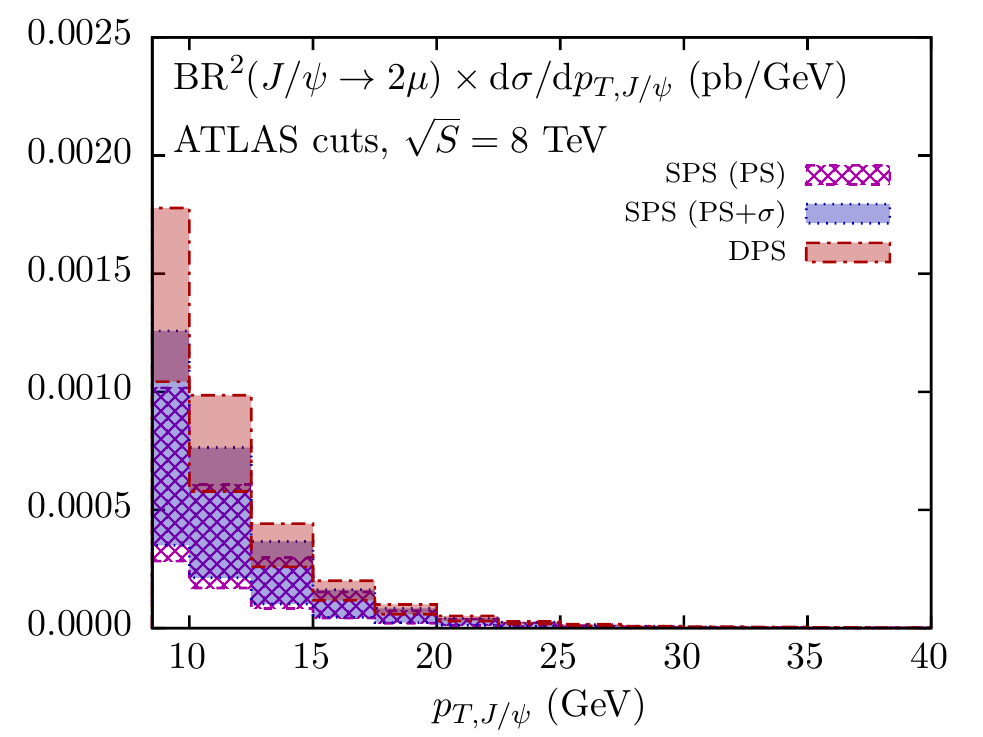}
	\end{tabular}
	\caption{The distribution showing the transverse momentum of a single $\jpsi$ for the ATLAS cuts at a collider energy of $\sqrt{S} = 8$ TeV.}
	\label{fig:atlas_pTjpsi}
\end{figure}
In Fig.~\ref{fig:atlas_plots}, the same set of distributions as in the case for LHCb is shown. The transverse momentum of the di-$\jpsi$ system is shown in Fig.~\ref{fig:atlas_plots}~(a). In the structure of the SPS distributions, one large peak at $p_{T,2\jpsi} \approx 20$~GeV is visible, while the distribution is relatively flat for $2.5~\mathrm{GeV} < p_{T,2\jpsi} < 10~\mathrm{GeV}$ compared to the DPS distribution. This behaviour for SPS can be understood by looking at the azimuthal angular separation, Fig.~\ref{fig:atlas_plots}~(b), where the SPS distributions favour two regions due to the recoil of the additional gluon radiation from parton showers: a forward-scattering region slightly below $|\Delta\phi_{2\jpsi}| = \frac{\pi}{2}$, and, to a lesser extent, the region towards back-to-back scattering with $\frac{\pi}{2} < |\Delta\phi_{2\jpsi}| \leq \pi$. Since the $p_T$ cut on the $\jpsi$ mesons requires them to have a transverse momentum of at least 8.5 GeV, and the $p_T$ distribution of a single $\jpsi$ is peaked towards the lowest possible values, see Fig.~\ref{fig:atlas_pTjpsi}, the back-to-back configuration mainly contributes to the flat region of the SPS distribution. The forward-scattering configuration does not play a role for low $p_{T,2\jpsi}$, but it contributes to the peak, approximately at twice the $p_T$ cut on the $\jpsi$. We have checked that the position of the peak is a consequence of the $p_T$ cut on the $\jpsi$ and moves when the cut is varied. The DPS signal does not show the same structure as for SPS, as its azimuthal angular separation is again uniformly distributed. Fig.~\ref{fig:atlas_plots}~(c) shows the invariant mass distribution of the di-$\jpsi$ system. A major difference to the LHCb cuts is that apart from a dominance of the DPS contributions at a high invariant mass, the distribution is now peaked at a much higher value of $M_{2\jpsi}$, allowing for a low-mass tail which is also dominated by DPS. The rapidity separation distribution in Fig.~\ref{fig:atlas_plots}~(d) shows a similar behaviour as for the LHCb case, indicating the possibility for a DPS signal measurement at high rapidity separations.

% #######################
\subsubsection{Predictions at 13 TeV}
% #######################
\begin{figure}[h!]
	\centering
	\begin{tabular}{cc}
		\includegraphics[width=0.47\columnwidth]{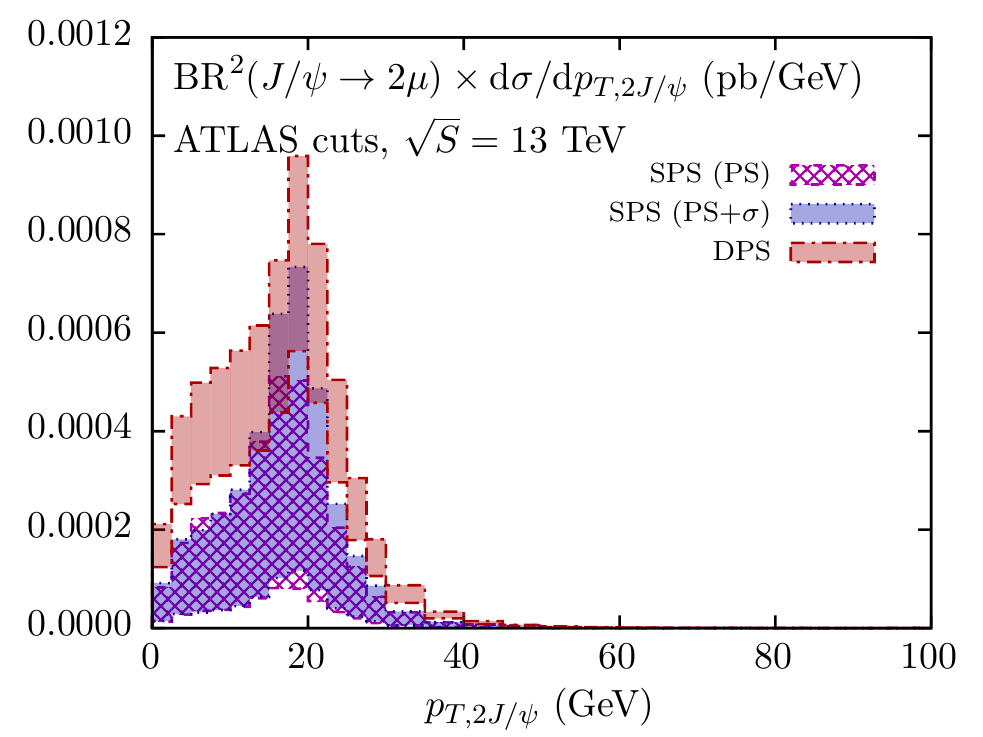}&
		\includegraphics[width=0.47\columnwidth]{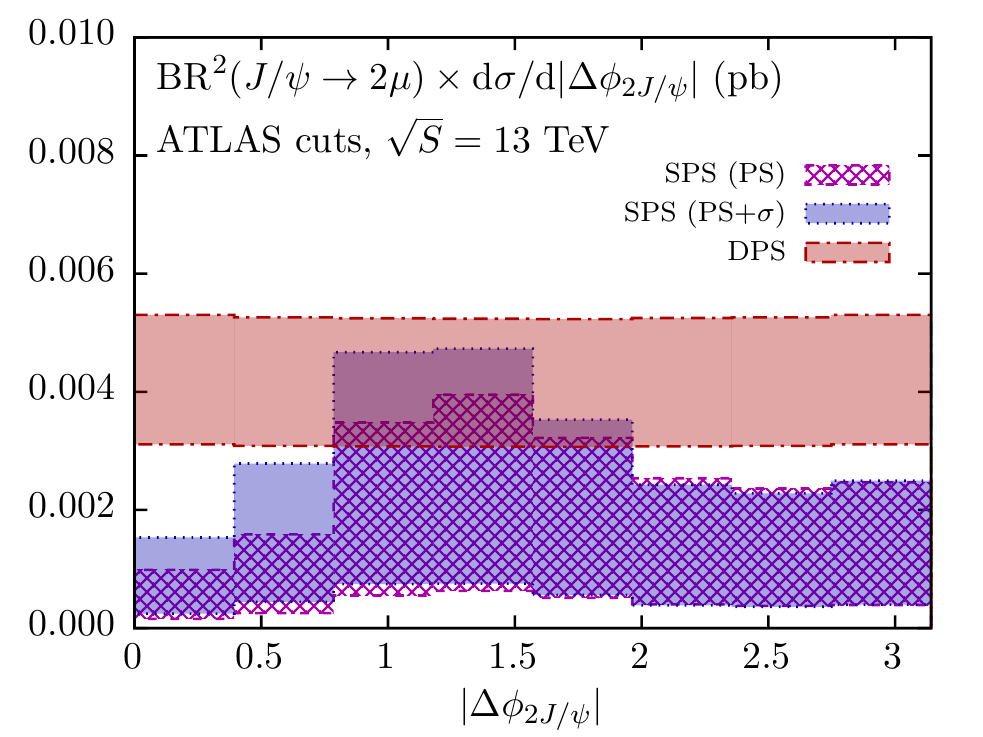}\\
		(a) & (b)\\
		\includegraphics[width=0.47\columnwidth]{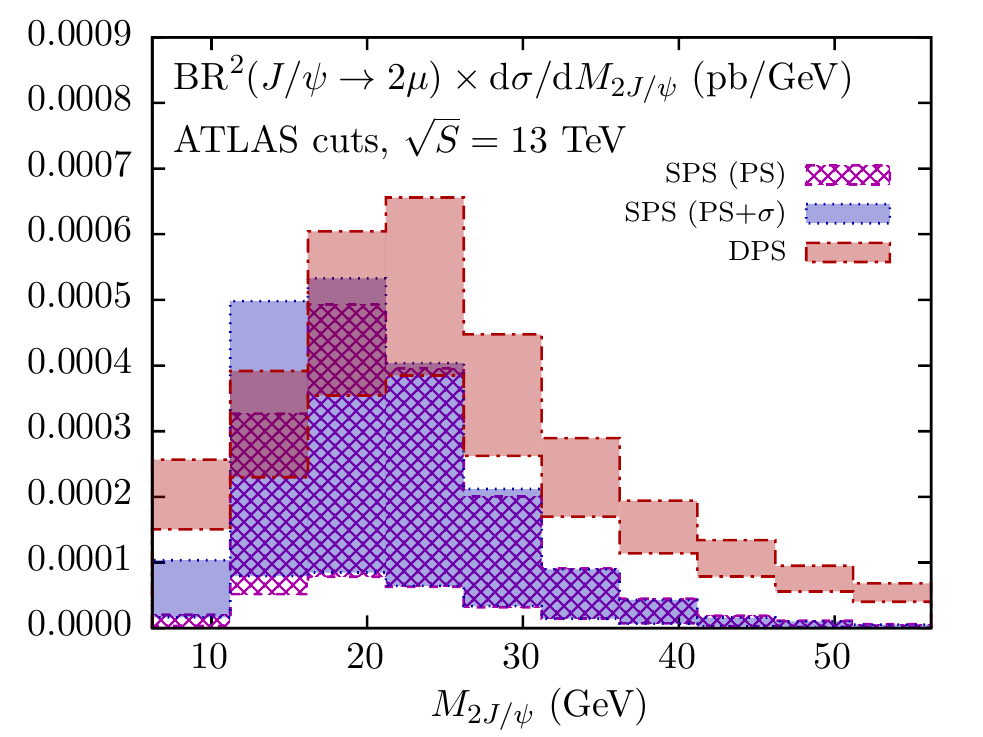}&
		\includegraphics[width=0.47\columnwidth]{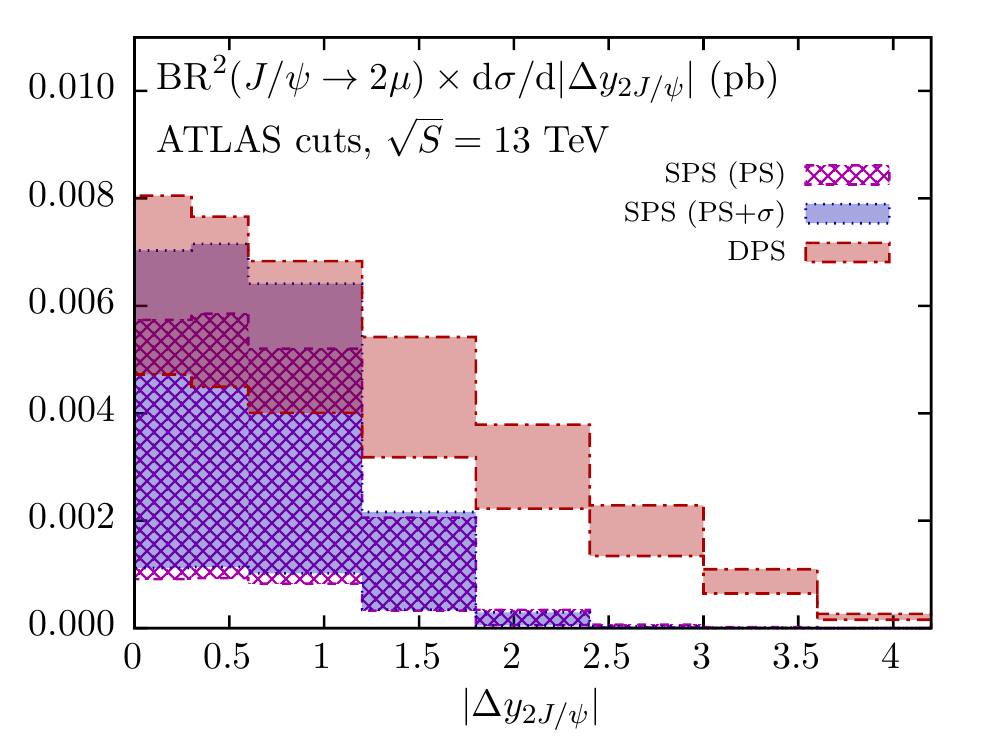}\\
		(c) & (d)
	\end{tabular}
	\caption{Differential distributions for the ATLAS cuts at a collider energy of $\sqrt{S} = 13$ TeV. The distributions are the same as for Fig.~\ref{fig:lhcb_plots}.}
	\label{fig:atlas_plots13}
\end{figure}
Also for the ATLAS predictions, shown in Fig.~\ref{fig:atlas_plots13}, the dominance of the DPS contributions at 13 TeV leads to an easier distinction between SPS and DPS. Despite the transverse momentum distribution of the di-$\jpsi$ system in Fig.~\ref{fig:atlas_plots13}~(a) now offering a clearer possibility to separate SPS and DPS contributions for low $p_{T,2\jpsi}$ due to the DPS contributions being larger than SPS by almost a factor of 2, we see that the increase in the centre-of-mass energy complicates the distinction of the shapes of SPS and DPS, and we again remark that higher-order corrections which are not included here can change the shape of the distribution significantly.

% #######################
\subsection{Comparison to CMS measurement}
% #######################
\begin{figure}[b!]
	\centering
	\begin{tabular}{cc}
		\includegraphics[width=0.47\columnwidth]{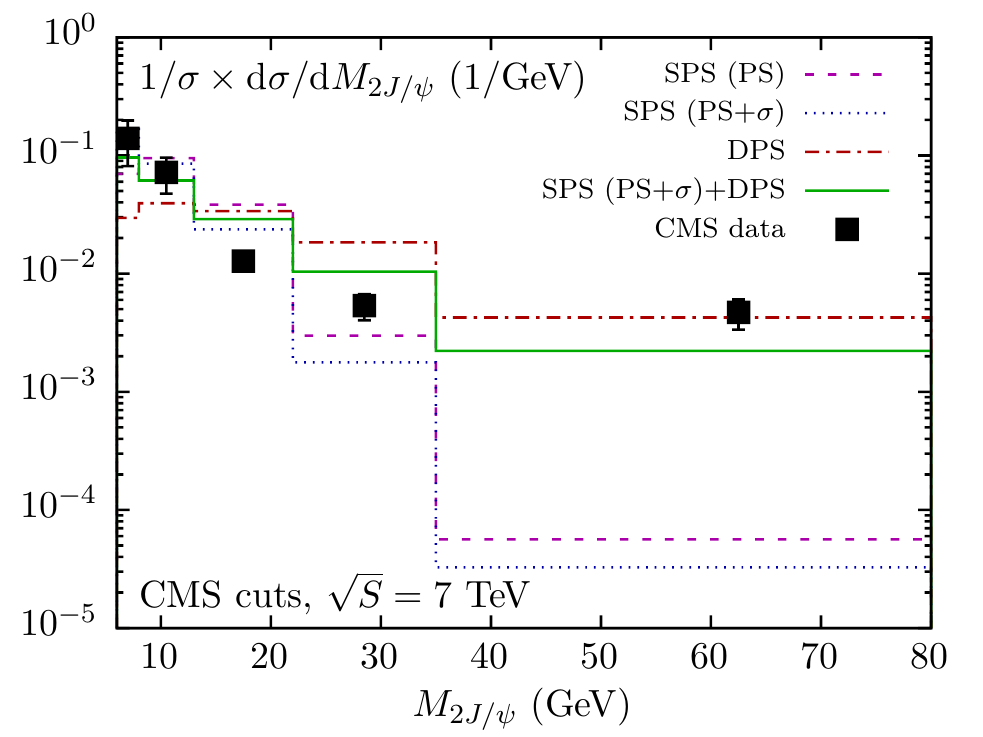}&
		\includegraphics[width=0.47\columnwidth]{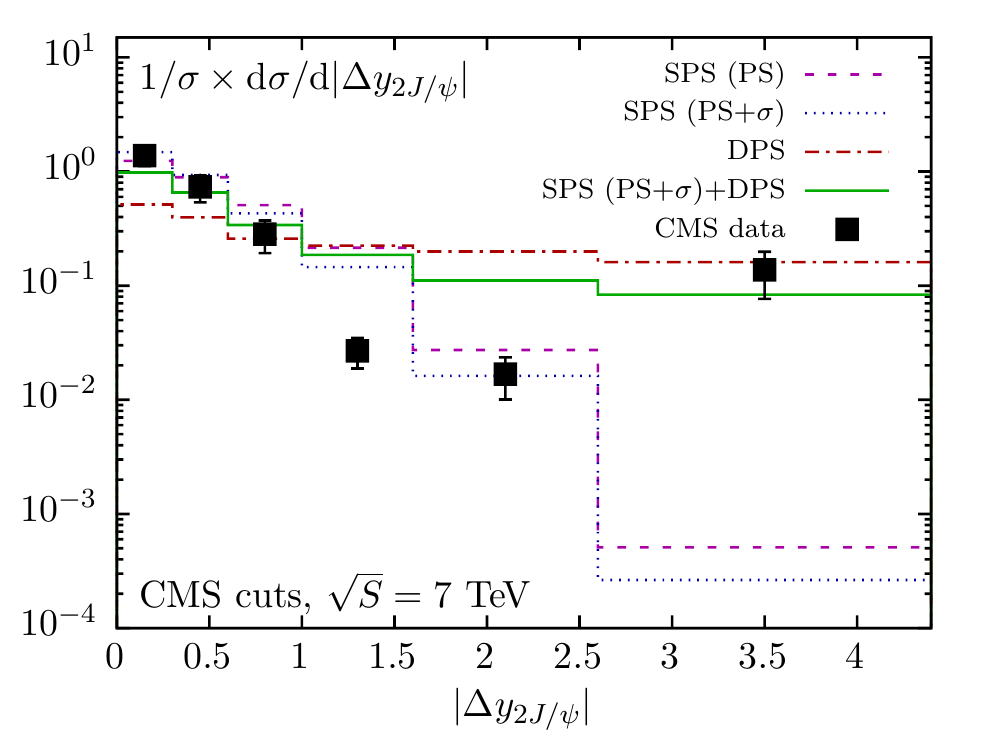}\\
		(a) & (b)\\
		\multicolumn{2}{c}{\includegraphics[width=0.47\columnwidth]{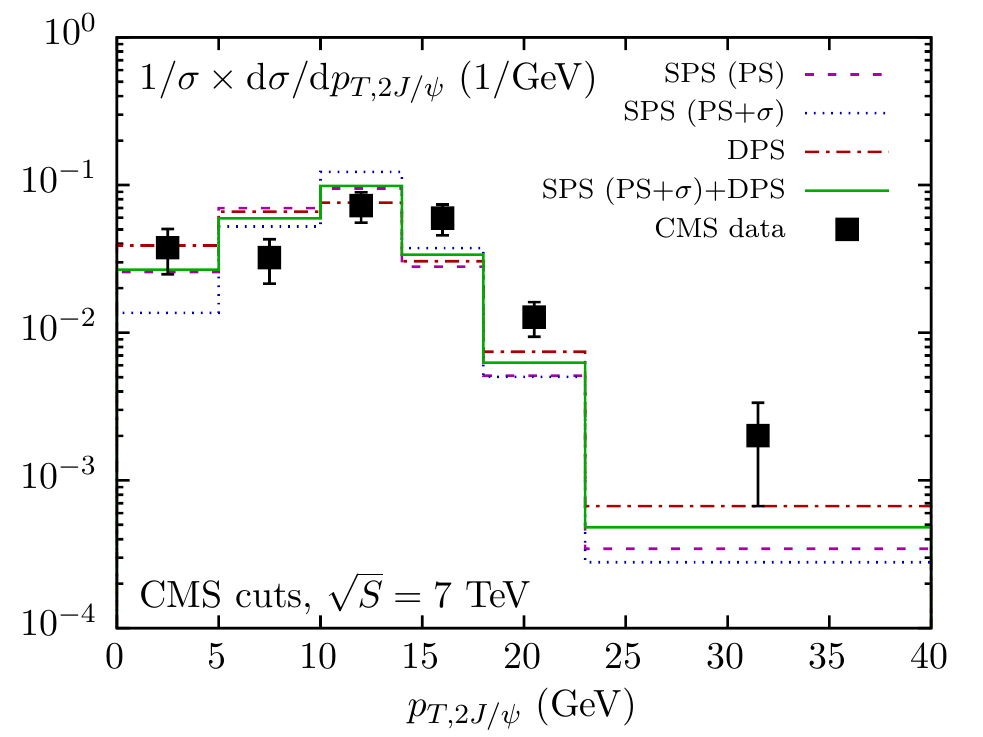}}\\
		\multicolumn{2}{c}{(c)}
	\end{tabular}
	\caption{Comparison of SPS and DPS predictions to the CMS data at a collider energy of $\sqrt{S} = 7$ TeV for the invariant mass (a), the rapidity separation (b), and the transverse momentum of the di-$\jpsi$ system (c). Shown are bins which are normalised to the corresponding total cross sections of the different SPS and DPS distributions and the data.}
	\label{fig:CMS_comp}
\end{figure}
The CMS experiment has recently measured $\jpsi$-pair production at $\sqrt{S}=7$~TeV \cite{Khachatryan:2014iia}. They applied the following cuts to their data:
\begin{itemize}
	\item[1)] $p_{T,\jpsi} > 6.5$ GeV for $|y_\jpsi| < 1.2$,
	\item[2)] $p_{T,\jpsi} > 6.5\to 4.5$ GeV for $1.2 < |y_\jpsi| < 1.43$,
	\item[3)] $p_{T,\jpsi} > 4.5$ GeV for $1.43 < |y_\jpsi| < 2.2$.
\end{itemize}
The $p_T$ cut in point 2 scales linearly from 6.5 GeV to 4.5 GeV with the value of $|y_\jpsi|$ from 1.2 to 1.43. No further cuts on the muons are applied.

In Fig.~\ref{fig:CMS_comp}, we compare our predictions to the CMS data. We show all bins normalised to the corresponding total cross section of a line to only compare the shape of the distributions and approximately remove the dependence on a specific PDF set. Furthermore, for the theoretical SPS and DPS predictions, we only show the central values without the error bands as described in section~\ref{sub:lhcb_predictions}. Shown are the invariant mass distribution, the transverse momentum of the di-$\jpsi$ system, and the rapidity separation of the two $\jpsi$. We see that our predictions catch the bulk behaviour of the CMS data, in particular also when further sources of uncertainty like the exact choice of the parameters which appear in the SPS and DPS calculations would be taken into account. Especially for the invariant mass and the rapidity separation distributions, Figs.~\ref{fig:CMS_comp}~(a) and (b), we see that at the high end of the spectrum the DPS contributions cannot be neglected. At the same time, the existing discepancies between theory and data call for further improvements in the theoretical description of SPS and DPS distributions.

% #######################
\subsection{Comparison to results of Lansberg and Shao}
\label{sec:complansberg}
% #######################

At last, we compare our results to the recently published ones of Lansberg and Shao \cite{Lansberg:2014swa}. The authors present predictions for similar scenarios of $\jpsi$-pair production at the LHCb and ATLAS experiments, with the difference of using full calculations of real gluon emission at NLO$^*$, the asterisk denoting the lack of virtual corrections. This method differs from ours by also taking into account hard gluon emission, while parton showering only considers soft gluons, however to all orders in $\alphas$. In this regard, it is interesting to see how the two approaches compare for the LHCb and ATLAS cuts. In order to minimise the sources of uncertainty, we choose parameters and PDF sets as close as possible to Lansberg et al. These are the wave function of the $\jpsi$ meson at origin $|R(0)|^2 = 0.81$ GeV$^3$, the charm mass in the range $m_c = 1.4$-$1.6$~GeV, and the PDF sets CTEQ6L1 for LO \cite{Pumplin:2002vw}, CTEQ6M for SPS (PS+$\sigma$) \cite{Pumplin:2002vw}, and MSTW2008 NLO for DPS. The renormalisation and factorisation scales for SPS production are set to $\mu_R = \mu_F = m^{\psi\psi}_T = \sqrt{(2m_\jpsi)^2+p_T^2}$, where $m_\jpsi = 2m_c$. The error bands are obtained from a simultaneous variation of $m_c$ and $\mu_R = \mu_F$ as $(1.4~\text{GeV},\frac{1}{2}m^{\psi\psi}_T)$;$(1.5~\text{GeV},m^{\psi\psi}_T)$;$(1.6~\text{GeV},2m^{\psi\psi}_T)$. For DPS, we additionally use the effective DPS cross section $\sigma_{\mathrm{eff}} = 8.2 \pm 2.2$~mb and the best fit parameters of \cite{Lansberg:2014swa} with $\kappa = 0.65$ and $\lambda = 0.32$. The factorisation scale in this case is set, as before, to the transverse mass of a single $\jpsi$. The error bands are now obtained from the uncertainty of $\sigma_{\mathrm{eff}}$.

\begin{figure}[t]
	\centering
	\begin{tabular}{cc}
		\includegraphics[width=0.47\columnwidth]{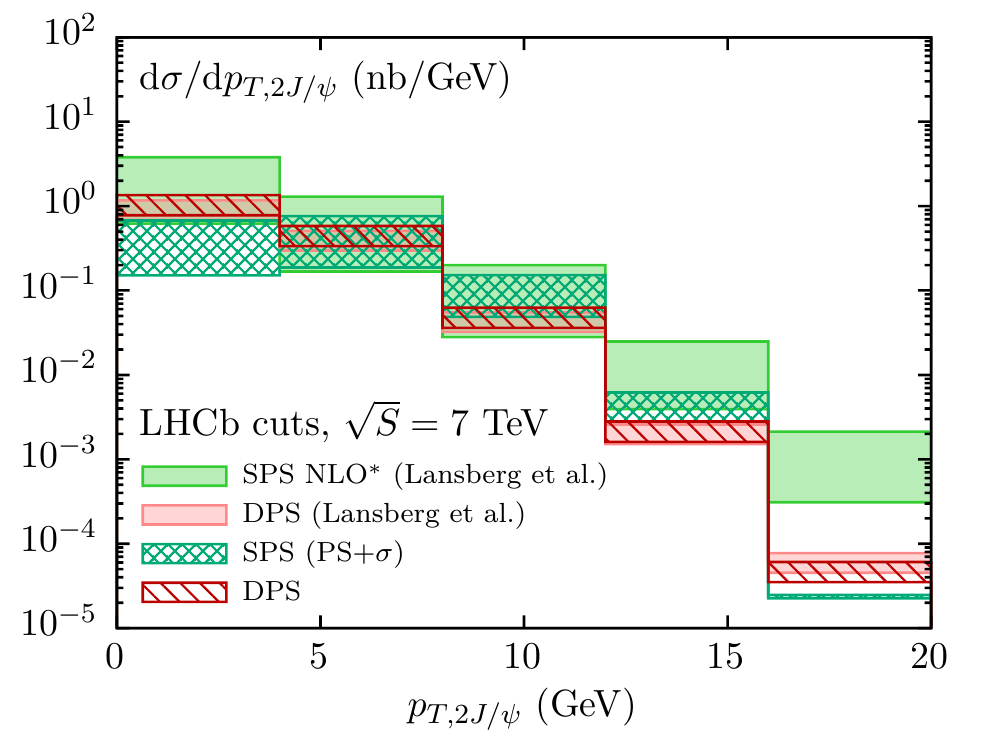}&
		\includegraphics[width=0.47\columnwidth]{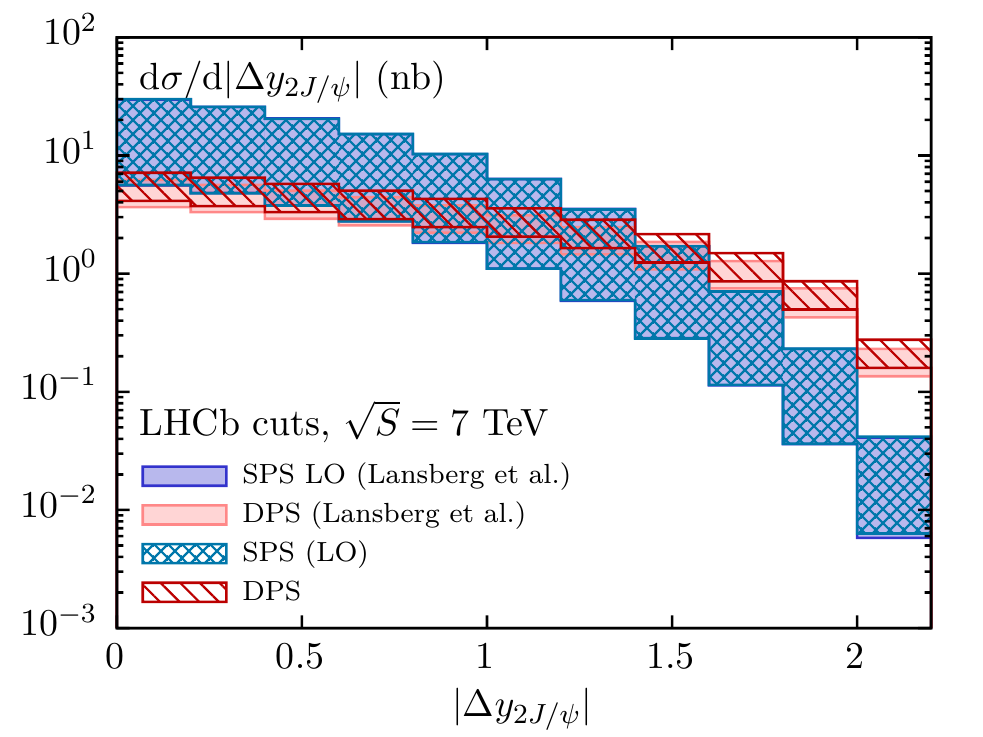}\\
		(a) & (b)\\
		\multicolumn{2}{c}{\includegraphics[width=0.47\columnwidth]{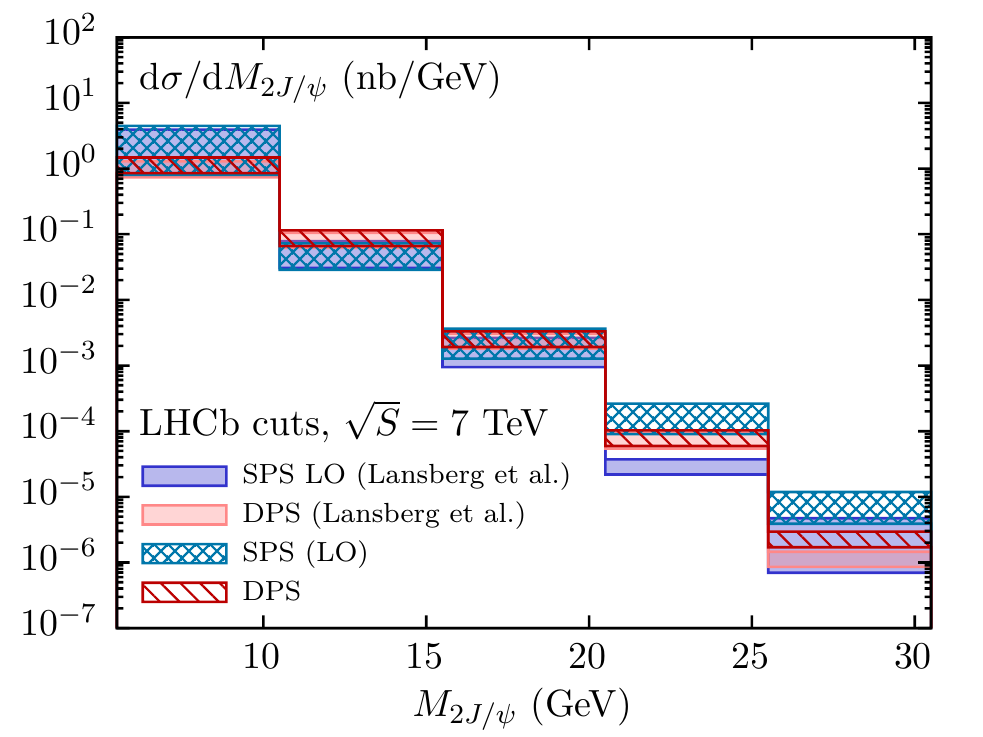}}\\
		\multicolumn{2}{c}{(c)}
	\end{tabular}
	\caption{Comparison between our results and \cite{Lansberg:2014swa} for the LHCb cuts. Shown are the transverse momentum distribution of the di-$\jpsi$ system (a), the rapidity separation (b), and the invariant mass (c). The ``SPS NLO (Lansberg et al.)'' and ``DPS (Lansberg et al.)'' uncertainty bands are read off from the corresponding plots in \cite{Lansberg:2014swa}. The cross sections here are not multiplied by the squared branching ratio $\mathrm{BR}^2(\jpsi\to 2\mu)$.}
	\label{fig:comp_lhcb_Lansberg}
\end{figure}

Fig.~\ref{fig:comp_lhcb_Lansberg} shows the comparison of distributions for three kinematic variables: the transverse momentum of the di-$\jpsi$ system $p_{T,2\jpsi}$, the rapidity separation between the two $\jpsi$ $|\Delta y_{2\jpsi}|$, and the invariant mass of the di-$\jpsi$ system $M_{2\jpsi}$, for the LHCb cuts at a collider energy of $\sqrt{S}=7$~TeV. We note that in \cite{Lansberg:2014swa}, the SPS predictions for the $|\Delta y_{2\jpsi}|$ and $M_{2\jpsi}$ distributions are only given at LO without taking into account additional gluon radiation.

In Fig.~\ref{fig:comp_lhcb_Lansberg}~(a), it can be seen that the transverse momentum distributions of SPS in our calculation and the one of \cite{Lansberg:2014swa} agree within the error bands for intermediate values of $p_{T,2\jpsi}$, while they differ at low and high $p_{T,2\jpsi}$. We see at low $p_{T,2\jpsi}$ the typical suppression from the all-order structure of parton showering, while the NLO$^*$ prediction is growing towards small $p_{T,2\jpsi}$. We expect that the inclusion of parton showering describes the shape of the distribution at low $p_{T,2\jpsi}$ better than a fixed-order calculation at NLO$^*$, since a major part of the contribution in this region comes from soft-collinear gluon emission, which is approximately taken into account at all orders in $\alphas$ in the parton shower formalism. On the other hand, the high-$p_{T,2\jpsi}$ region cannot be described properly by parton showering due to the lack of hard gluon emission which dominates this region. We remark that the good agreement for intermediate $p_{T,2\jpsi}$ is also related to the $p_{T,\jpsi} < 10$~GeV cut, effectively cutting off the high-$p_{T,2\jpsi}$ region where hard gluon emission becomes important. The DPS predictions for the transverse momentum distribution agree very well between our calculation and \cite{Lansberg:2014swa} due to the same functional form of the cross section fit of Eq.~\eqref{crystalball}.
\begin{figure}[t]
	\centering
	\begin{tabular}{cc}
		\includegraphics[width=0.47\columnwidth]{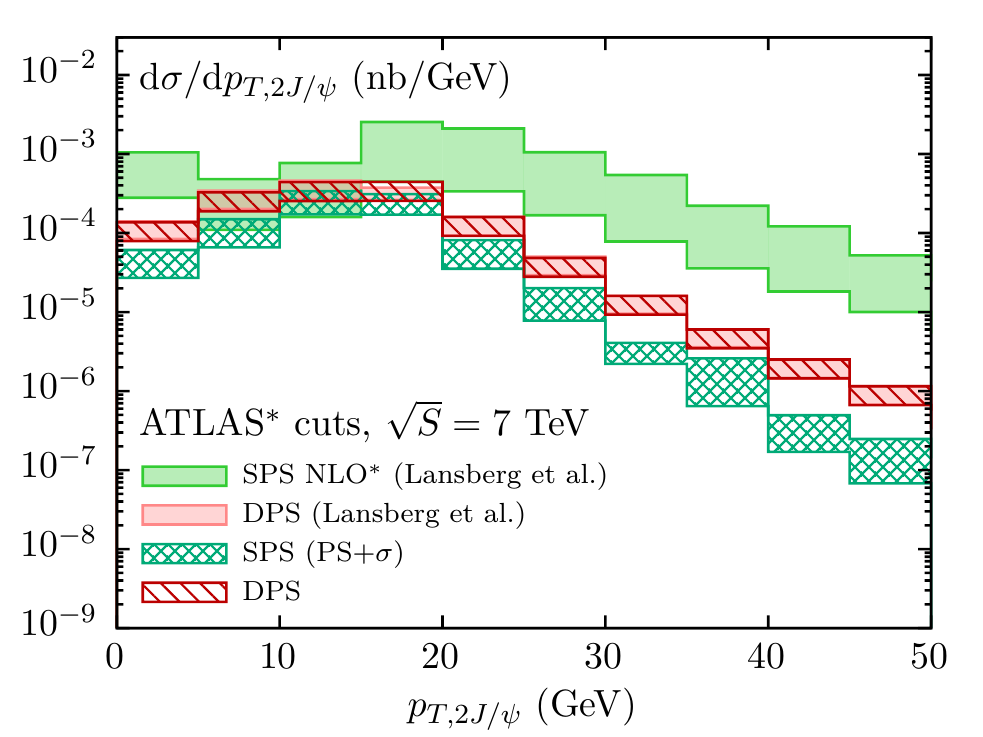}&
		\includegraphics[width=0.47\columnwidth]{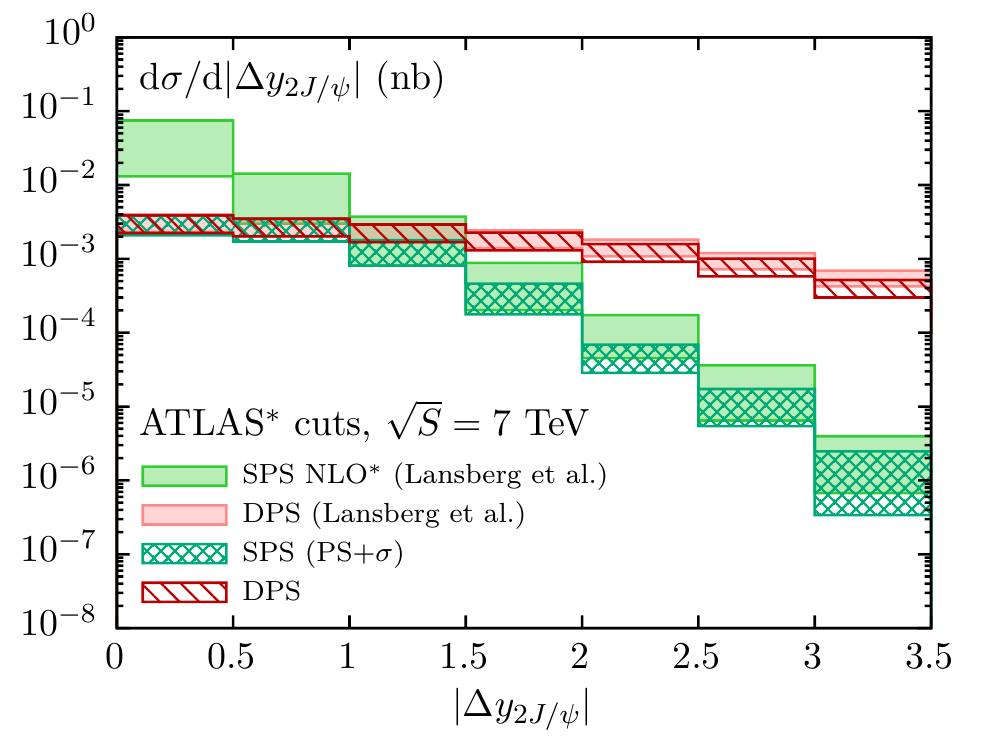}\\
		(a) & (b)\\
		\multicolumn{2}{c}{\includegraphics[width=0.47\columnwidth]{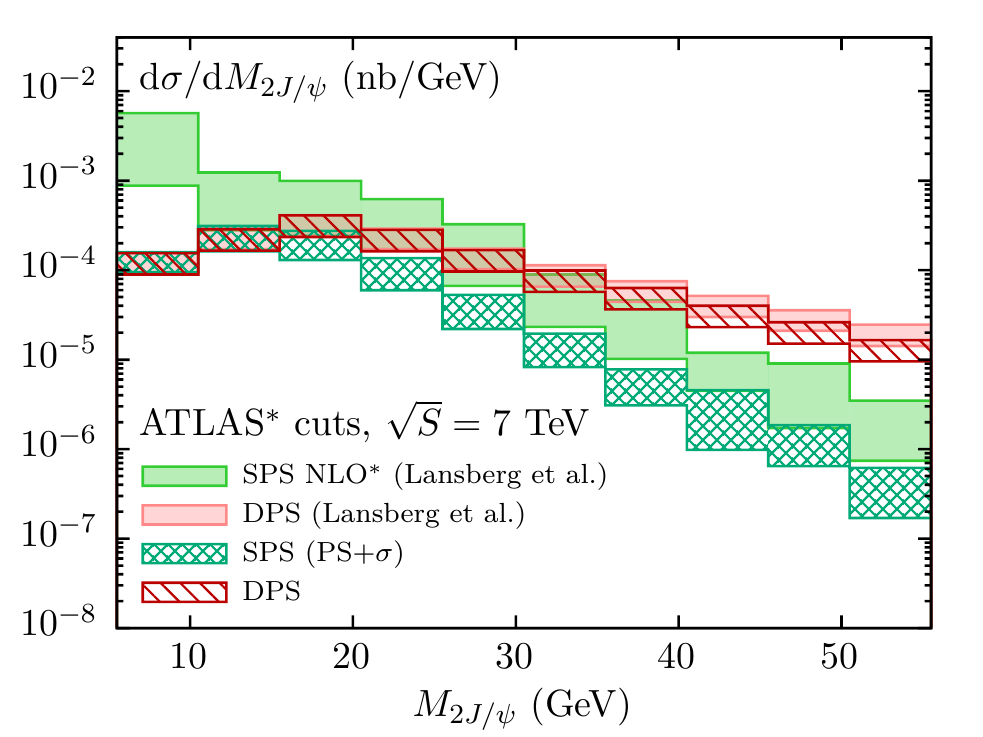}}\\
		\multicolumn{2}{c}{(c)}
	\end{tabular}
	\caption{Same as Fig.~\ref{fig:comp_lhcb_Lansberg} for the case of the ATLAS$^*$ cuts. These differ from the ATLAS cuts defined in section~\ref{subsec:atlascuts} by imposing a lower $p_{T,\jpsi}$ cut of $p_{T,\jpsi} > 5$~GeV instead of $p_{T,\jpsi} > 8.5$~GeV. Furthermore, the predictions are shown for $\sqrt{S} = 7$~TeV instead of 8 TeV.}
	\label{fig:comp_atlas_Lansberg}
\end{figure}
The rapidity separation in Fig.~\ref{fig:comp_lhcb_Lansberg}~(b) shows, as expected, a very good agreement between the SPS and DPS predictions from us and \cite{Lansberg:2014swa} because of the LO calculations and the same parametrisation of Eq.~\eqref{crystalball}. For the invariant mass distributions of Fig.~\ref{fig:comp_lhcb_Lansberg}~(c), our SPS and DPS predictions agree well with \cite{Lansberg:2014swa} for an invariant mass up to approx.\ 20 GeV, while there are differences for SPS in the last two bins and for DPS in the last bin. These differences might be related to numerical precision, as the differential cross sections become very small for a high invariant mass.

Fig.~\ref{fig:comp_atlas_Lansberg} for the ATLAS$^*$ predictions at a collider energy of $\sqrt{S}=7$~TeV shows the same set of distributions as for the LHCb predictions. It should be noted that the asterisk denotes a changed ATLAS cut with $p_{T,\jpsi} > 5$~GeV instead of $p_{T,\jpsi} > 8.5$~GeV.

The transverse momentum distribution of Fig.~\ref{fig:comp_atlas_Lansberg}~(a) displays a larger $p_{T,2\jpsi}$ range than for the LHCb cuts, which shows that for large values of $p_{T,2\jpsi} > 20$~GeV, the parton shower and NLO$^*$ results of SPS differ by a notable amount\footnote{Note that in any case calculations in the CSM cannot deliver an accurate theoretical description of the large $p_T$ region; instead the NRQCD framework should be used.} while there is again an agreement for low $p_{T,2\jpsi} \approx 5$-15~GeV within the error bands. We point out that the bulk of the cross section comes from the region for $p_{T,2\jpsi} < 20$~GeV, as seen e.g.\ in Fig.~\ref{fig:atlas_plots}~(b) for a similar setup at 8~TeV (on a linear axis), so that the differences for $p_{T,2\jpsi} > 20$~GeV between the parton shower and NLO$^*$ results affect the description of only a small portion of events. Interestingly, while there is a small difference for SPS in the bin with small rapidity separation, Fig.~\ref{fig:comp_atlas_Lansberg}~(b), the two predictions agree well within the errors for $|\Delta y_{2\jpsi}| > 0.5$. The invariant mass distribution of Fig.~\ref{fig:comp_atlas_Lansberg}~(c) shows that, while there is again a difference for the smallest bin, the predictions for SPS with parton shower and NLO$^*$ corrections almost agree within the error bands (and they in fact do for some bins), although it can be seen more clearly here that the lack of hard gluon emission leads to the parton shower result always being below the NLO$^*$ result. The DPS predictions agree very well for the transverse momentum distribution, while for the rapidity separation and the invariant mass, there are slight deviations in the high-$|\Delta y_{2\jpsi}|$ and $M_{2\jpsi}$ bins, possibly related to the difference in numerics and codes used to compute these predictions.

From these comparisons, we see that, as one would expect, the rapidity separation distribution is most stable with respect to higher-order corrections from hard gluon emission that are not included in our approach, while the transverse momentum distribution of the di-$\jpsi$ system is most strongly affected by them. We remark that here the SPS and DPS predictions have been computed with different input PDFs (CTEQ6L1, CTEQ6M, and MSTW2008 NLO, respectively) for the purpose of comparing to \cite{Lansberg:2014swa}, while the comparison between the magnitudes of SPS and DPS presented in section~\ref{sec:kin_dis} avoids introducing PDF effects unrelated to the SPS and DPS calculations.

% #######################
\section{Conclusions}\label{sec:conclusions}
% #######################
Precise predictions for multi-parton interations are a vital ingredient for the high-energy collisions at the LHC, in particular during the current run at a centre-of-mass energy of $\sqrt{S}=13$~TeV and future runs at higher energies, where the probability for such subleading scattering processes to happen is significantly increased. In this work we have documented SPS and DPS predictions for the production of $\jpsi$ pairs with the updated fiducial volume cuts for the LHCb analyses of Run I data, and also for a new DPS study of $\jpsi$-pair production at $\sqrt{S}=8$~TeV by the ATLAS experiment. The distributions show interesting indications that DPS processes could contribute significantly to certain kinematic regions of the invariant mass and rapidity separation of the di-$\jpsi$ system, while the transverse momentum distribution is very susceptible to higher-order corrections. The predictions for a collider energy of $\sqrt{S}=13$~TeV show a considerable increase of the DPS contributions with respect to SPS. Finally, the comparison to the results presented in \cite{Lansberg:2014swa} indicate a good agreement for regions where it is reasonable to compare a parton shower to a NLO$^*$ calculation, supporting the parton shower approach as a good approximation in these regions. \\

\noindent
\textbf{Note added:} In the final preparation stages of this report, we have become aware of a new ATLAS study of double $\jpsi$ production~\cite{ATLAS:2016eii}. In this study, our DPS predictions presented in section~\ref{sec:kin_dis} are compared with the data-driven estimates of DPS. A good agreement between DPS theory and data is found for all differential distributions reported in~\cite{ATLAS:2016eii}. The full $\jpsi$ distributions measured by ATLAS are then compared with the sum of our DPS predictions and NLO SPS predictions of~\cite{Lansberg:2014swa} with the collision energy and fiducial volume adjusted according to the experimental analysis and normalised to the fraction of DPS events found using data-driven model. We have checked that applying the same normalisation procedure to our predictions leads to a rather good agreement with the measured $\jpsi$ distributions, apart from the large end of the spectra (and the first, low end bins in some cases), in accordance with observations in~\cite{ATLAS:2016eii} and section~\ref{sec:complansberg}. It needs to be checked if supplementing the theoretical predictions with full NLO corrections can eliminate the need for introducing the normalisation procedure, as results of~\cite{Sun:2014gca} would suggest.

% #######################
\section*{Acknowledgements}
% #######################
The authors thank C.H.\ Kom for sharing his expertise during initial stages of the event simulation. Part of this work has been performed on the High Performance Computing cluster PALMA maintained by the Center for Information Technology (ZIV) at WWU Münster, and on the high-performance computing resources funded by the Ministry of Science, Research and the Arts and the Universities of the State of Baden-Württemberg, Germany, within the framework program bwHPC. CB acknowledges support by the Institutional Strategy of the University of Tübingen (DFG, ZUK 63).  AK would like to thank the Theory Group at CERN, where part of this work was carried out, for its kind hospitality.

% #######################

\providecommand{\href}[2]{#2}\begingroup\raggedright\endgroup

\end{document}